\documentclass[12pt]{iopart}
\pdfoutput=1
\usepackage{color}
\usepackage{amssymb}
\usepackage[export]{adjustbox}
\usepackage{ulem}
\usepackage{graphicx} 

\def\beqra{\begin{eqnarray}}
\def\eeqra{\end{eqnarray}}
\def\beq{\begin{equation}}
\def\eeq{\end{equation}}
\def\Hc{{\cal H}}

\def\etain{\eta_{in}}

\def\vp{\bar{\varphi}}

\def\vp{\varphi}

\def\bx{{\bf{x}}}
\def\bk{{\bf{k}}}
\def\bp{{\bf{p}}}
\def\bq{{\bf{q}}}

\def\bV0{{\bf{V_0}}}
\def\re#1{(\ref{#1})}

\def\agt{~\mbox{\raisebox{-.6ex}{$\stackrel{>}{\sim}$}}~}
\def\alt{~\mbox{\raisebox{-.6ex}{$\stackrel{<}{\sim}$}}~}
\def\bx{{\bf{x}}}
\def\by{{\bf{y}}}
\def\br{{\bf{r}}}

\def\bk{{\bf{k}}}
\def\bp{{\bf{p}}}
\def\bq{{\bf{q}}}

\def\vp{\varphi}

\begin{document}

\title[Asymptotic expansions for the Large Scale Structure]{Asymptotic expansions for the Large Scale Structure}

\author{Shi-Fan Chen$^a$ and Massimo Pietroni$^{b,c}$}
\vskip 0.3 cm
\address{
$^a$Department of Physics, University of California, Berkeley, CA 94720, USA\\
$^b$INFN, Sezione di Padova, via Marzolo 8, I-35131, Padova, Italy\\
$^c$Dipartimento di Scienze Matematiche, Fisiche ed Informatiche dell'Universit\`a di Parma, Parco Area delle Scienze 7/a, I-43124, Parma, Italy
}

\begin{abstract}
We explore the deep ultraviolet (that is, short-distance) limit of the power spectrum (PS) and of the correlation function of a cold dark matter dominated Universe. While for large scales the PS can be written as a double series expansion, in powers of the linear PS and of the wavenumber $k$, we show that, in the opposite limit, it can be expressed via an expansion in  powers of the form $1/k^{d+2n}$, where $d$ is the number of spatial dimensions, and $n$ is a non negative integer. The coefficients of the  terms of the  expansion  are nonperturbative in the linear PS, and can be interpreted in terms of the probability density function for the displacement field, evaluated around specific configurations of the latter, that we identify.  In the case of the Zel'dovich dynamics, these coefficients can be determined analytically, whereas for the exact dynamics they can be treated as fit, or nuisance, parameters.
We confirm our  findings with numerical simulations and discuss the necessary steps to match our results to those obtained for larger scales and to actual measurements.
\end{abstract}

\maketitle
\section{Introduction}
Analytical approaches to the evolution of the Large Scale Structure (LSS) \cite{PT} represent a complementary tool with respect to N-Body simulations, their main virtues being flexibility (namely the possibility to apply such methods to a wide range of cosmologies beyond standard $\Lambda$CDM) and computational speed. Both these aspects make these approaches essential tools for the efficient exploration of cosmological parameter space in the analysis of data from present and future surveys. The canonical example of such approaches is Standard Perturbation Theory (SPT), which treats structure formation within the framework of Eulerian fluid dynamics organized order-by-order in the linear power spectrum (PS), though alternative approaches within the Lagrangian framework (LPT) similarly organizing the statistics of fluid displacements are also popular. The main limitation of these approaches is the incompleteness and non-convergence of the perturbative expansion they are based on (see, for instance, \cite{Blas:2013aba}), which limits their range of applicability to very large scales. Standard perturbation theory, for example, is not able to provide percent level evaluations of the matter power spectrum for wavenumbers $k\agt 0.05 \,{\rm h/Mpc}$ at $z=0$. More recently, these issues have led to a re-interpretation of cosmological perturbation theory within the language of effective field theory, wherein the effects of short-distance, or ultraviolet (UV), physics are tamed and parametrized through counterterms that can be fit to simulated or observed data  \cite{Baumann:2010tm, Pietroni:2011iz, Carrasco:2012cv, Manzotti:2014loa, McQuinn:2015tva, Vlah:2015sea, Noda:2017tfh}. These methods yield a derivative expansion governed by the nonlinear scale $k_{\rm nl}$, and can extend the reach of perturbation theory up to $k\sim 0.15 \,{\rm h/Mpc}$ at $z=0$ (and $k\sim 0.4  \,{\rm h/Mpc}$ at $z=1$), though necessarily become invalid past $k_{\rm nl}$ where the UV physics can no longer be described by a finite set of parameters and the series in $k/k_{\rm nl}$ is non-convergent.

The ultimate reason for the failure of the SPT-based approaches even for relatively simple collisionless cold dark matter (CDM) dynamics is physical: at short scales where the velocity dispersion and vorticity play increasingly relevant roles, CDM cannot be described as a perfect fluid. In particular, the former implies  the  breakdown of the ``single stream'' approximation SPT is based on. The emergence of multistreaming, also known as shell-crossing, is marked by the generation of higher order moments of the distribution function (beyond density and velocity) and by singularities in the mapping between the initial and final positions of the fluid elements (`Lagrangian' to `Eulerian' mapping). These mathematical difficulties have long been considered as insurmountable obstacles to the continuation of analytical methods beyond shell-crossing. 

In recent years, however, a number of authors have reconsidered these difficulties, making substantial progress in our understanding of physics beyond the single-stream regime. Most of this attention has been devoted to describe the transition from single stream to multistreaming (the ``first shell-crossing"), see for instance \cite{Taruya:2017ohk, Rampf:2017jan, McDonald:2017ths, Saga:2018nud, Halle:2020zln}. In \cite{Pietroni:2018ebj} we showed explicitly (in 1 spatial dimensions) the emergence of nonperturbative terms as the effect of shell-crossing. By nonperturbative we mean here terms that vanish faster than any positive power law in the linear PS as the latter goes to zero. These terms are also generically non-analytical, that is, they include logarithms or fractional powers of the PS, which, as was argued in  \cite{Pietroni:2018ebj}, signals the emergence of non-locality in Lagrangian space. These new terms become more and more relevant with increasing time and density contrasts and cure the pathological behavior of the asymptotic SPT expansion. 
Moreover, it was also shown that dynamics post shell crossing is governed by attractors which make the mapping between Lagrangian and Eulerian space flatter and flatter inside multistreaming regions.  

In this paper we pursue our exploration of the dynamics well beyond shell-crossing by studying the limits of very small scales  ($r\to 0$) in the correlation function and of asymptotically large wavenumbers ($k\to \infty$) in the power spectrum (PS). In other terms, we embark on a journey to the antipodes of the SPT domain. 

We will frame our exploration of the deep ultraviolet regime of structure formation in the Lagrangian formulation of fluid mechanics, which relates the initial (Lagrangian) to the final (Eulerian) coordinate of a given CDM particle and is exact for CDM systems if stream crossing is taken into account. Our work is not the first to explore the ultraviolet regime of structure formation within the Lagrangian picture (see e.g. \cite{Valageas:2010rx, Valageas:2010yw, Seljak:2015rea}). 
However, unlike in previous work we will express the PS and the correlation function as a path integral over displacement field configurations, weighted by a probability density functional (pdf), in a form which is particularly suited for taking the asymptotic limits. 

Our main result is an  expansion in powers of $1/k$ for the PS at large $k$'s. In 1 + 1 dimensions, the leading term scales as $1/k$ with a coefficient given by the pdf for the first derivative of the displacement field, $\psi^{(1)}(q)$, evaluated at  $\psi^{(1)}(q)=-1$. This quantity is intrinsically non-perturbative: for instance, in Zel'dovich dynamics it can be computed analytically and  is proportional to the non-perturbative quantity $e^{-1/2 \sigma_\delta^2}$, where $\sigma_\delta^2$ is the variance of the linear density contrast. Moreover, Lagrangian coordinates $q$  with $\psi^{(1)}(q)=-1$ correspond precisely to points where shell-crossing is happening. Therefore, the abundance of these points is what governs the large scale behavior of the PS. Finally, as discussed in \cite{Pietroni:2018ebj} and recalled above, in exact 1+1 dynamics, $\psi^{(1)}(q)=-1$ acts as a late-time attractor, such that the $1/k$ asymptote is reached for $k$'s smaller than in the case of Zel'dovich dynamics, where this attractor is not present. 

Moving from 1+1 dimensions to 3+1 dimensions, we find analogous behavior. The asymptotic behavior of the PS is now $1/k^3$, both in Zel'dovich and in real dynamics, and the relevant field configurations are a straightforward three-dimensional counterpart of those found in one dimension.

Strictly within the realm of CDM, our results provide constraints on the very small scale behavior of the PS which should be satisfied by any other approach aiming at describing the LSS at small scales. In particular, we will discuss how the halo model seems to be in tension with the requirements found in this paper, and therefore needs to be corrected at very small scales.

The paper is organized as follows. In Sect.~\ref{SPT} we discuss SPT at the field and at the correlation function levels in 1+1 dimensions, reviewing its breakdown and the emergence of nonperturbative behavior. Staying in the one-dimensional world, in Sect.~\ref{PSS}  we derive our main result, namely, the asymptotic expansion for the PS both for the full dynamics and for the Zel'dovich one, and discuss the dependence of our results on the smoothing scale. We extend our results to three dimensions in Sect.~\ref{3d}, and discuss the relations of our results to the halo model in Sect.~\ref{halolink}. We conclude in Sect.~\ref{fine}. Finally, in \ref{npZeld} we give analytical results for the correlation function in Zel'dovich dynamics, and in \ref{numerics} we describe in detail the numerical solutions to the 1+1 dimensional field equations.

\section{SPT and its Failure in 1+1 Dimensions}
\label{SPT}
We begin with the dynamics of collisionlesss dark matter in one spatial dimension, using it as an illustrative example of the failures of traditional perturbation theory techniques. Related discussions can be found in \cite{McQuinn:2015tva,Pietroni:2018ebj}.

The (Eulerian) position of a point-like matter element in one dimensions is given by
\beq
x(q,\tau) = q+\psi(q,\tau)\,,
\eeq
where $q$ denotes the uniformly-distributed initial (Lagrangian) coordinate and the displacement $\psi(q,\tau)$ solves the equation of motion
\beq
\ddot\psi(q,\tau) +{\cal H} \dot\psi(q,\tau) = F(q,\tau)\,.
\label{fulleq}
\eeq
Dots indicate derivatives with respect to conformal time $\tau$,  ${\cal H}=\dot a/a$ is the conformal Hubble parameter and $a(\tau)$ the scale factor, normalized to one at present time $\tau_0$. Using the well-known fact that uniform sheets of matter produce uniform gravitational fields on either side, the force $F(q,\tau)$ is given by \cite{Pietroni:2018ebj}
\beqra
&&\!\!\!\!\!\!\!\!\!\!\!\!\!\!\! F(q,\tau)=-\frac{3}{2}{\cal H}^2 \int dq'\left[ \Theta(q+\psi(q,\tau) -q'+\psi(q',\tau)) - \Theta(q+\psi(q,\tau) -q')  \right]\\
&& \;\,= \frac{3}{2}{\cal H}^2 \sum_{i=1}^{N_s(x(q,\tau),\tau)} (-1)^{i+1} \psi(q_i,\tau)\,,
\label{force}
\eeqra
where $\Theta(x)$ is the Heaviside's function, and the sum in the second line is taken over all the $N_s(x(q,\tau),\tau)$ streams, namely, the $q_i$'s solving the equation 
\beq
q_i+\psi(q_i,\tau)=x(q,\tau)\,.
\label{roots}
\eeq
Of course, one of the $q_i$'s is $q$ itself; if it is also the only solution, then $q$ is a ``single-stream'' point, otherwise it is a ``multistreaming'' one. 

In the following, we will consider both the solution of the `full' dynamical equation, Eq.~\re{fulleq}, and the solution of the  equation for the Zel'dovich dynamics, which is obtained by setting
\beq
 F(q,\tau) \to  F^Z(q,\tau)=\frac{3}{2}{\cal H}^2 \psi(q,\tau)\,.
 \label{ZeldF}
\eeq
The Zel'dovich dynamical equation can be solved analytically for an Einstein-de Sitter universe, to get the two independent solutions
\beq
\psi^Z(q,\tau) =a(\tau)^m\psi^Z(q,\tau_{0}) \,,
\label{zeldg}
\eeq
where  $m=\,1,\,-3/2$ for the growing and the decaying mode, respectively. In the following we will consider initial conditions containing only the growing mode, which is selected by setting
\beq
\dot\psi(q,\tau_{in})= {\cal H}(\tau_{in}) \psi(q,\tau_{in})\,.
\label{vin}
\eeq
Notice that, for single stream points,  Eq.~\re{force}  gives $ F(q,\tau)=\frac{3}{2}{\cal H}^2 \psi(q,\tau)$, and, therefore,  Zel'dovich dynamics is exact.  On the other hand, in 3+1 dimensions, Zel'dovich dynamics is never exact. 

\subsection{Field level}

We want to investigate the conditions for the validity of a perturbative expansion  at the field level, that is, before computing statistics such as correlation functions. We start from the exact relation giving the density contrast in Eulerian space in terms of the displacement field,
\beq
1+\delta(x,\tau)= \int dq\,\delta_D(q+\psi(q,\tau)-x) = \sum_{i=1}^{N_s(x,\tau)}\frac{1}{\left| 1+\psi^{(1)}(q_i(x,\tau),\tau)\right|}\,,
\label{true}
\eeq
where $\psi^{(n)}(q,\tau)$ denotes the $n$-th derivative of $\psi(q,\tau)$ with respect to $q$.

The perturbative expansion can be obtained by  functionally expanding in $\psi(q,\tau)$ the expression containing the integral of the delta function,
\beqra
&&\delta(x,\tau) \simeq \int dq\,  \sum_{n=1}^\infty \frac{\psi(q,\tau)^n}{n!}\frac{\partial^n}{\partial q^n} \delta_D(q-x)\,, \nonumber\\
&& \qquad\;\;=  \sum_{n=1}^\infty \frac{(-1)^n}{n!} \frac{\partial^n}{\partial x^n} \int dq\, \psi(q,\tau)^n \delta_D(q-x)\,, \nonumber\\
&&\qquad\;\;=  \sum_{n=1}^\infty \frac{(-1)^n}{n!} \frac{\partial^n}{\partial x^n}\psi(x,\tau)^n\,,
\eeqra
where the symbol ``$\simeq$'' here indicates that the two quantities are equal if the  series converges.

Without loss of generality, let us set $x = 0$. The regularity of $\psi(x,\tau)$ implies that it can be expanded around $x=0$ as
\beq
\psi(x,\tau) = b_0(\tau)+b_1(\tau) x+b_2(\tau) x^2+b_3(\tau) x^3+ \cdots\,.
\eeq
 The constant term, $b_0(\tau)$ can be set to zero by a change of frame, therefore, the SPT expansion of the density contrast in $x=0$ is, simply,
\beq
 \delta(x,\tau) = \sum_{n=1}^\infty (-b_1(\tau))^n\simeq - \frac{b_1(\tau)}{1+b_1(\tau)} =- \frac{\psi^{(1)}(x=0,\tau)}{1+\psi^{(1)}(x=0,\tau)}  \,.
\label{SPTconv}
\eeq
Convergence then requires $|\psi^{(1)}(x=0,\tau)|<1$, while the requirement that the last expression makes sense as a density contrast (that is, that $-1\le \delta(x=0,\tau) <\infty$) corresponds to $\psi^{(1)}(x=0,\tau)> -1$. Convergence, however, is not enough. The SPT series should converge to the true answer (see Eq.~\re{true}), which now reads,  
\beq
\delta(x=0,\tau)=  \sum_{i=1}^{N_s(x=0,\tau)}\frac{1}{\left| 1+\psi^{(1)}(q_i(x=0,\tau),\tau)\right|} -1\,.
\label{true2}
\eeq
We see immediately that, if the point under consideration is a multistreaming one, that is, if $N_s(x=0,\tau)>1$, the SPT series either does not converge or converges to the wrong answer.  

If $x=0$ is a single stream point ($N_s(x=0,\tau)=1$ and $q(x=0,\tau)=0$) then the RHS's of  Eqs.~\re{SPTconv} and \re{true2} coincide, and therefore if  the SPT series converges, it does so to the true answer.   Since  $\psi^\prime(q_i(x=0,\tau),\tau)$ scales as $a(\tau)$,  at a sufficiently early time its modulus is less than unity and the series converges.  At later times, SPT can break down for two reasons depending on the sign of $\psi^\prime(q_i(x=0,\tau),\tau)$.  If it is positive and breaks the convergence barrier from below, it means that the region around $x=0$ is becoming emptier and emptier. The limiting value $\psi^\prime(q_i(x=0,\tau),\tau)=1$  corresponds to a nonlinear density contrast $\delta(x=0,\tau)=-1/2$, a value for which nothing dramatic happens. In this regime, one can extend the SPT result \re{SPTconv} beyond the convergence limit and still get the right answer up to  $\psi^\prime(q_i(x=0,\tau),\tau)\to +\infty$, corresponding to the empty limit ($\delta(x=0,\tau)\to -1$). This also suggests a nonlinear redefinition of the field as, for instance \cite{Pajer:2017ulp},
\beq
\lambda(x=0,\tau) = -\log\left(1+\psi^\prime(q_i(x=0,\tau),\tau) \right)\,,
\eeq
so that the density field is given by
\beq
\delta(x=0,\tau) = e^{\lambda(x=0,\tau) } -1\,,
\eeq
and can be expanded perturbatively in $\lambda(x=0,\tau)$ over the whole range $-\infty< \lambda(x=0,\tau)<+\infty$ corresponding to the enlarged range $-1<\psi^\prime(q_i(x=0,\tau),\tau)<+\infty$.

On the other hand, when $\psi^\prime(q_i(x=0,\tau),\tau)$ reaches the value $-1$ (from above) in a single-stream region, something dramatic happens: the density contrast \re{true} diverges, marking the transition into the multistreaming regime. The divergence can be regularized  by smoothing the density contrast in Eulerian space, as discussed in \cite{Pietroni:2018ebj}. After smoothing, the divergence is traded for non-analytic behavior in terms of the smoothing length. This signals the onset of non-locality in Lagrangian space, namely the fact that different Lagrangian regions contribute to the density contrast in a given Eulerian point. 

As we will see, the special value $\psi^\prime(q_i(x=0,\tau),\tau)=-1$ plays a crucial role in the determination of the asymptotic small scale behavior of the PS and of the correlation function.

\subsection{Correlation function}
\label{cfss}
The correlation function is given by
\beqra
&&1+\xi(r)=1+\langle\delta\left(\frac{r}{2}\right) \delta\left(-\frac{r}{2}\right) \rangle= \langle  \int_{-\infty}^{\infty} dq\,\delta_D\left(r-q-\Delta\psi\left(q\right) \right)
\rangle\label{2}\\
&&\,\;\;\qquad\;\;= \langle\sum_{i=1}^{N(r,\tau)} \frac{1}{\left| 1+\Delta\psi^{(1)}(q_i(r))\right|} \rangle ,
\label{3}
\eeqra
where
\beq
\Delta\psi\left(q\right) \equiv \psi\left(\frac{q}{2}\right) - \psi\left(-\frac{q}{2}\right) \,,
\label{Dpsi}
\eeq
the roots $q_i(r)$ are the solutions of 
\beq
q+\Delta \psi(q) = r\,,
\eeq
and we have omitted the time dependence. The brackets in the above expressions indicate the average over configurations of the relative displacement field $\Delta\psi\left(q \right)$, weighted with a given probability, evolved from the initial one through the field equations, Eq.~\re{fulleq}.
Notice that, for any fixed $r \neq 0$, the values of the roots $q_i(r)$ in general differ from configuration to configuration,  whereas, for $r=0$, $q=0$ is always a solution, since $\Delta\psi\left(0\right)=0$ (see Eq.~\re{Dpsi}). Therefore, when computing the correlation function in $r=0$, we expect a logarithmic divergent contribution, 
\beq
  \xi(0)\ni \langle  \frac{1}{\left| 1+\Delta\psi^{(1)}(0)\right|}   \rangle \sim \int d \Delta\psi^{(1)}(0) {\cal P}[ \Delta\psi^{(1)}(0)]  \frac{1}{\left| 1+\Delta\psi^{(1)}(0)\right|} \,,
  \label{expdiv}
\eeq
where $ {\cal P}[ \Delta\psi^{(1)}(0)]$ is the probability density function of $ \Delta\psi^{(1)}(0)$. On the other hand, for $r\neq 0$ the divergence is smoothed by the stochasticity induced by the fact that the corresponding roots $q_i(r)$ depend on the  field configurations one is averaging over. Therefore, we anticipate that $\xi(r)$ diverges logarithmically in $r$ as $r \to 0$ regardless of the considered dynamics, as long as $ {\cal P}[  \Delta\psi^{(1)}(0)=-1]\neq 0$. In \ref{npZeld} we will show it explicitly in the case of the Zel'dovich dynamics, while in the next section we will see how the pdf in $ \Delta\psi^{(1)}(0)=-1$ is related to the large $k$ limit of the PS.

The correlation function  can always be split in a perturbative and a nonperturbative part, 
\beqra
&& \!\!\!\!\!\!\!\!\!\!\!\!\!\!\! \!\!\!\!\!\!\!\! \!\!\!\!\!\!\!\! \xi(r)=\xi_{\rm{pert}}(r)+\xi_{\rm{nonpert}}(r)\nonumber\\
&& \!\!\!\!\!\!\!\!\!\!\!\! \!\!\!\!\!\!\!\!=\langle \sum_{n=1}^{\infty} \frac{(-1)^n}{n!} \frac{\partial^n}{\partial r^n}\Delta \psi(r)^n  \rangle_{\rm single} +
\langle\sum_{i=1}^{N(r)} \frac{1}{\left| 1+\Delta\psi^{(1)}(q_i(r))\right|} -1 \rangle_{\rm multi} \, \label{spt1} \\
&& \!\!\!\!\!\!\!\!\!\!\!\! \!\!\!\!\!\!\!\!= \langle\frac{- \Delta\psi^{(1)}(q(r))   }{1+ \Delta\psi^{(1)}(q(r)) }  \rangle_{\rm single} +
\langle\sum_{i=1}^{N(r)} \frac{1}{\left| 1+\Delta\psi^{(1)}(q_i(r))\right|} -1 \rangle_{\rm multi} \,,
\eeqra
where the first average is taken over field configurations such that $r$ is a single stream point, that is, $N(r)=1$, whereas the second one is taken over all the remaining ones (that is, $N(r)\ge 3$), such that $ \langle \cdots \rangle_{\rm single}+
\langle \cdots \rangle_{\rm multi} = \langle \cdots \rangle$. On single stream points the perturbative expansion  is guaranteed to converge  (see Eq.~\re{3})  to the first term at the last line. From the splitting above we see that a nonperturbative contribution is generally  present, even at very early times, unless the probability of getting multistreaming at $r$ is exactly zero. Of course, the larger $r$ and the smaller $\tau$, the more negligible we expect these nonperturbative contributions to be. On the other hand,  the perturbative series 
$\sum_{n=1}^{\infty} (-1)^n (\partial^n/\partial r^n)\Delta \psi(r)^n/n!$, evaluated on multistreaming configurations, is  non-convergent, and this is the reason of the inevitable failure of the SPT expansion.

\section{Asymptotic Behavior of the Power Spectrum}
\subsection{Formalism}
\label{PSS}

We will now derive the main result of our paper in 1+1 dimensions, i.e. the asymptotic small-scale behavior of the power spectrum. Taking the Fourier transform of the correlation function in Eq.~\re{2} gives the well-known result for the PS,
\beq
P(k)+ 2 \pi\,\delta_D(k)= \langle\int_{-\infty}^{+\infty} dq \, e^{i k \, q} e^{i k \,\Delta \psi(q)}\rangle\,.
\label{psd}
\eeq
The ensemble average above can be expressed as a path integral
\beq
P(k)+2 \pi\,\delta_D(k)= \frac{1}{k}\int {\cal D} \psi \, {\cal P}[\psi] \,{\cal F}[\psi](k) \,,
\label{path}
\eeq
where the functional $ {\cal P}[\psi] $ weights each displacement field configuration with its proper probability, while the functional ${\cal F}[\psi](k) $ can be read from \re{psd},
\beq
\!\!\!\!\!\!\!\! {\cal F}[\psi](k) \equiv k\int dq\;  e^{i k \, q} e^{i k \,\Delta \psi(q)}\,.
\label{funcF}
\eeq

One way to express the path integral explicitly is to compactify the $q$-domain on a line of length $L$ and  Fourier expand the relative displacement field as
\beq
\Delta\psi(q)=\psi\left(\frac{q}{2} \right)-\psi\left(-\frac{q}{2} \right)=\frac{4}{L}\sum_{n=0}^{N_{\rm{max}}} |\tilde \psi_n | \sin \vp_n\,\sin\left(\frac{2\pi n q}{L}   \right)\,,
\label{Fpsi}
\eeq
where the Fourier amplitudes $|\tilde \psi_n |$ and the phases $\vp_n$ are real numbers in the domains $[0,+\infty)$ and $[0,2\pi)$, respectively, with $N_{\rm{max}}$ related to the spatial resolution in Lagrangian space  $\Delta q$ through $L/N_{\rm{max}}=\Delta q$.
The functional ${\cal F}[\psi](k)$ can now be expressed as
\footnote{The $1/2$ coefficient in front of the integral, as well as the integration domain from $-L$ to $L$, keep track of the fact that $\Delta \psi(q)$ is defined for $q\in[-L,L)$, 
while $\psi(q)$ is defined for $q\in[-L/2,L/2)$, and the Fourier transforms are taken on the same interval. In other terms, for vanishing displacement field, one has to recover $F_L[\{|\tilde 
\psi_n |=0\},\{\vp_n\}](k)/k=2 \pi \delta_D(k)$, with the delta function normalized as $2 \pi \delta_D(k=0)=L$.}
\beq
\!\!\!\!\!\!\!\! {\cal F}[\psi](k) \to F_L[\{|\tilde \psi_n |\},\{\vp_n\}](k)\equiv \frac{k}{2} \int_{-L}^{L} dq \,e^{i kq} e^{i \frac{4}{L}\sum_{n=0}^\infty |\tilde \psi_n | \sin \vp_n\,\sin\left(\frac{2\pi n q}{L}   \right)}\,.
\eeq
To proceed in the computation we need the joint pdf of the amplitudes and phases,
\beq
\int {\cal D} \psi \, {\cal P}[\psi]\to \Pi_{n} \left(\int_0^\infty d |\tilde \psi_n | \int_0^{2 \pi} d \vp_n \right) {\cal P}[\{|\tilde \psi_n |\},\{\vp_n\}]\,.
\eeq
In the case of the Zel'dovich dynamics, each mode follows an independent Rayleigh distribution,
\beq
{\cal P}_Z[\{|\tilde \psi_n |\},\{\vp_n\}]=\Pi_n\, \frac{1}{2\pi}\, \frac{|\tilde \psi_n |}{\sigma_n^2} e^{-\frac{|\tilde \psi_n |^2}{2\sigma_n^2}}\,,
\label{incond}
\eeq
where 
\beq
\frac{\sigma_n^2}{L}= \frac{1}{2}\frac{P_{\rm{lin}}(p_n)}{p_n^2}\,, \qquad\qquad\left(p_n\equiv \frac{2n\pi}{L}\right)\,.
\label{incond2}
\eeq
In this case, the path integral can be performed analytically, giving the well known result
\beqra
&&\!\!\!\!\!\!\!\!\! \!\!\!\!\!\!\!\!\! \!\!\!\!\!\!\!\!\! P_Z(k) +(2\pi)\delta_D(k)= \frac{1}{2} \int_{-L}^{L}dq\, e^{i k q}e^{-\frac{1}{2} \left( \frac{4 k }{L}\right)^2 \sum_{n=0}^\infty \sigma_n^2 \sin^2\left(\frac{\pi n q}{L}\right) }  \nonumber\\
&&\quad\quad\;\;\;\;\;\to \int_{-\infty}^{+\infty} dq\, e^{i k q} e^{-k^2\sigma_{\Delta\psi}^2(q)}\,,
\eeqra
where we have taken the $L\to \infty$ limit and $\sigma_{\Delta\psi}^2(q)$ is defined  in \re{sigdeltapsi}. One can check that the above PS is the Fourier transform of \re{xiZ}.

However, our main focus in this paper is on the behavior at large $k$, where the functional ${\cal F}[\psi](k) $ can be more conveniently expressed in an alternative way. To see this, let us Taylor expand the displacement field, expressing  $\Delta \psi(q)$  as
\beq
\Delta \psi(q)\simeq \vp_1 \, q + \frac{\vp_3 }{24} \,q^3 + \cdots +\frac{\vp_{2n+1}}{2^{2n} (2n+1)!} \,q^{2n+1}+\cdots \,,
\eeq
where $\vp_n$ indicates the n$^{\rm th}$ derivative of the displacement field evaluated at  $q=0$, $\vp_n=\psi^{(n)}(0)$. Note that the $\vp_n$'s are real numbers and not functions.
We now have
\beqra
&&\!\!\! {\cal F}[\psi](k) \to G[\{ \vp_{2n+1}\}](k) \equiv   k \int_{-\infty}^{\infty} dq\, e^{ikq}e^{ik \sum_{n=0}^\infty \frac{\vp_{2n+1}}{2^{2n} (2n+1)!} q^{2n+1}}\,,\nonumber\\
&&\quad\quad\qquad\qquad\qquad\qquad\,= \int_{-\infty}^{\infty} dy\, e^{iy}e^{i \sum_{n=0}^\infty \frac{\vp_{2n+1}}{  2^{2n} (2n+1)!} \frac{y^{2n+1}}{k^{2n}}}\,,
\label{Gf}
\eeqra
and, correspondingly, 
\beq
\int {\cal D} \psi \, {\cal P}[\psi]\to  \left(\Pi_{n=0}^\infty \int d   \vp_{2n+1}\right) {\cal P}[\{ \vp_{2n+1}\}]\,.
\eeq
\begin{figure}[t]
\centering 
\includegraphics[width=.45\textwidth,clip]{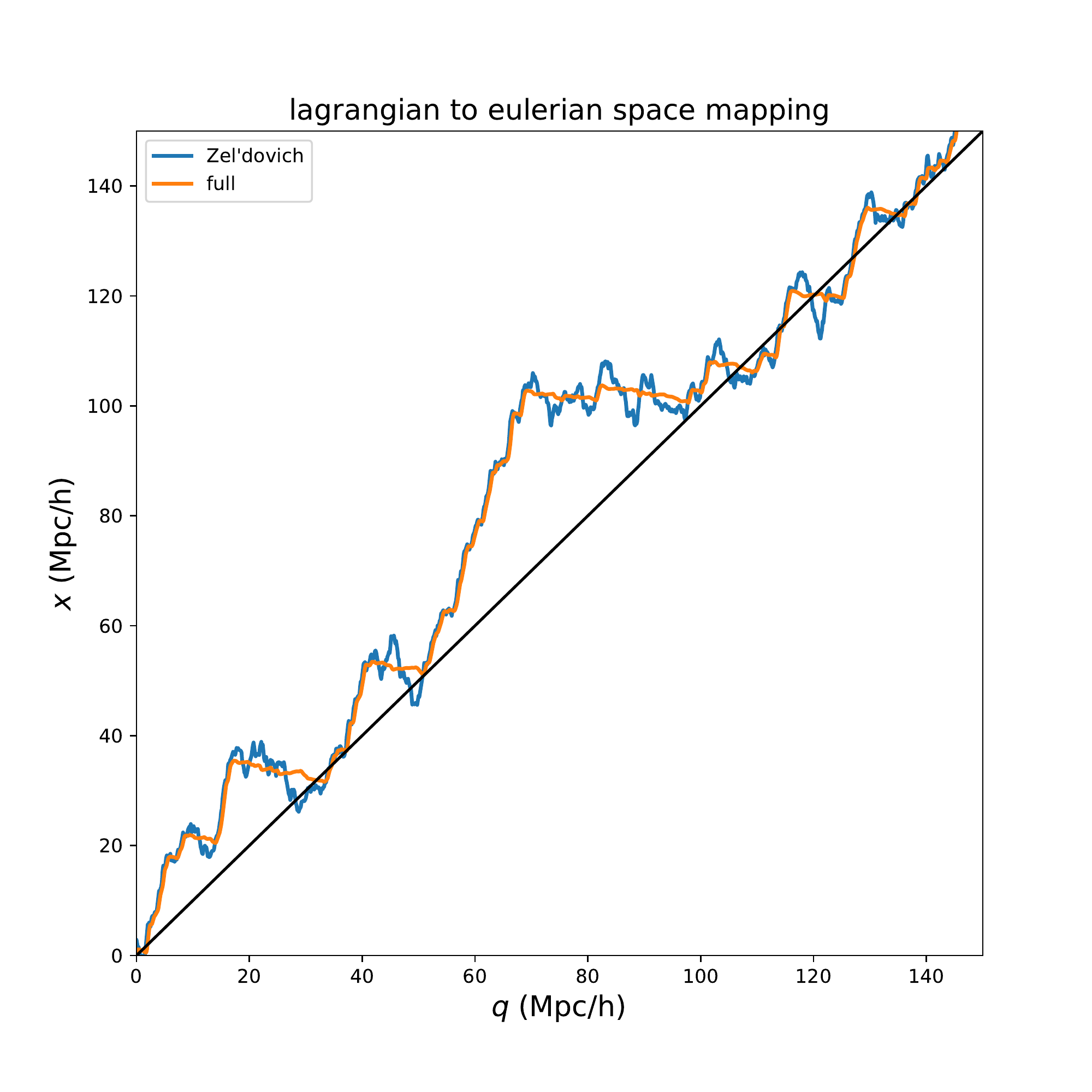}
\includegraphics[width=.45\textwidth,clip]{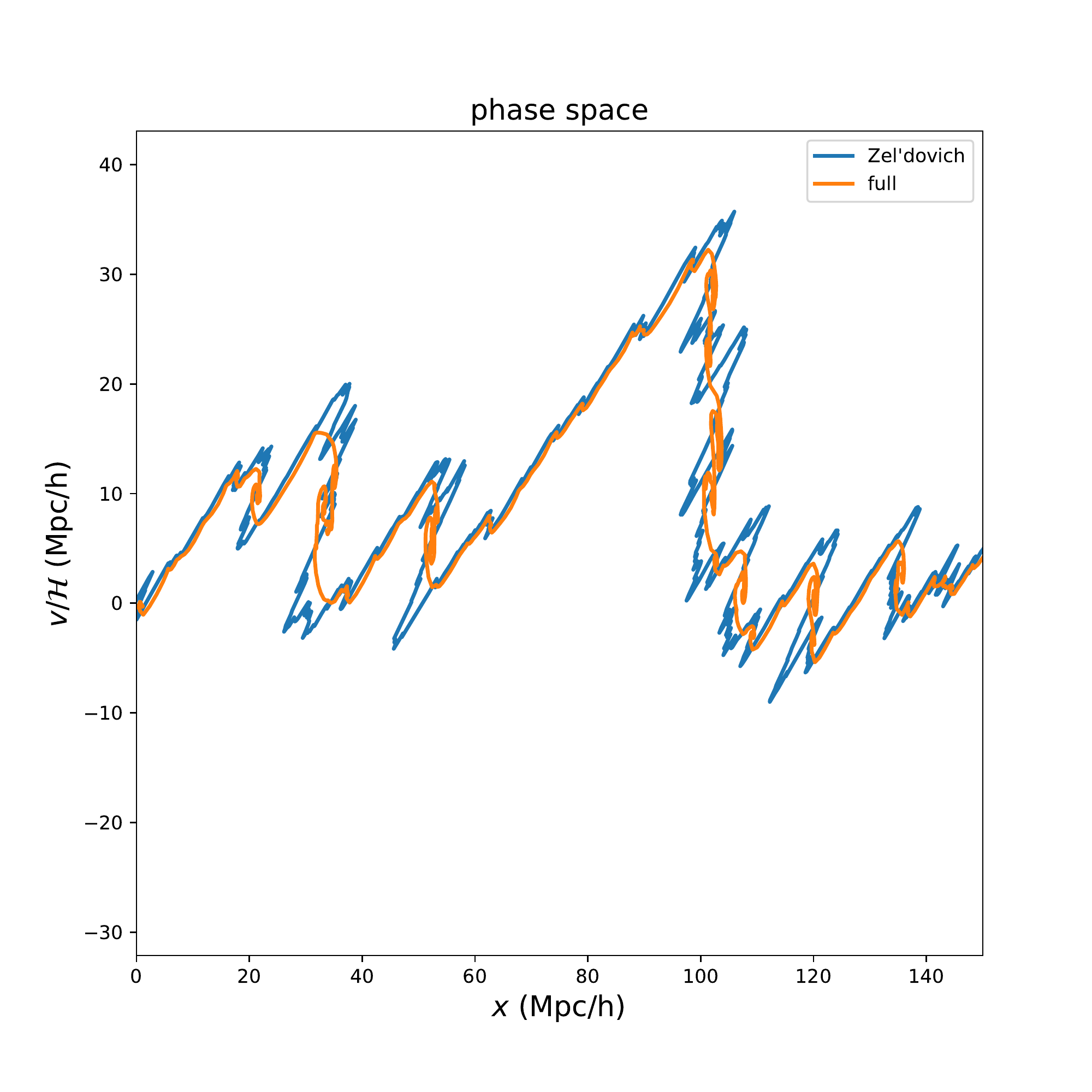}
\caption{Left: the mapping between Lagrangian coordinate $q$ and the Eulerian one, $x$, at $z=0$ in a portion of our simulation.  Right: phase space diagram in Eulerian space. In both panels, orange is for the full dynamics, blue for the Zel'dovich one.  }
\label{numsol}
\end{figure}

 The small scale limit of the PS is then governed by the statistical averages of the odd derivatives of $\psi(q)$, with  derivatives of higher orders being suppressed by higher orders in $1/k$. The  $k \to \infty$ limit is controlled by the $n=0$ term in the sum at the exponent,  
 \beq
 \lim_{k\to \infty}  G[\{ \vp_{2n+1}\};k] = \int_{-\infty}^{+\infty} e^{i y \left(1+\vp_1\right)} = 2\pi \delta_D  \left(1+\vp_1\right)\,.
 \eeq
 The path integral can then be computed in this limit as
 \beqra
&&\lim_{k\to \infty} P(k) =  \left(\Pi_{n=0}^\infty \int d   \vp_{2n+1}\right) {\cal P}[\{ \vp_{2n+1}\}]  \frac{2\pi}{k}\delta_D  \left(1+\vp_{1}\right)\,,\nonumber\\
&&\qquad\quad\;\;\;\;= \frac{2\pi}{k} {\cal P}_{ \psi^{(1)}}[-1]\,,
\label{C0}
 \eeqra
 where $ {\cal P}_{ \psi^{(1)}}[\vp_1] $ is the probability density function (pdf) for $ \psi^{(1)}$
 \beq
 {\cal P}_{ \psi^{(1)}}[ \vp_1]\equiv \left(\Pi_{n=1}^\infty \int d   \vp_{2n+1}\right)\, {\cal P}[\{ \vp_{2n+1}\}]\,.
 \eeq
 
The subleading terms in the large $k$ limit can be derived by expanding the exponential in \re{Gf}, to obtain a series in inverse powers of $k$,
\beq
P(k) \sim \frac{{\cal C}_0}{k}+\frac{{\cal C}_1}{k^3}+\frac{{\cal C}_2}{k^5}+\cdots+\frac{{\cal C}_n}{k^{2n+1}}\,,\qquad\qquad({\rm for}\;\; k\to\infty)\,.
\label{expas}
\eeq
From \re{C0} we already  have
\beq
{\cal C}_0 =\lim_{k\to\infty} k P(k)=\lim_{k\to\infty}  \langle {\cal F}[\psi](k)\rangle = 2\pi\, {\cal P}_{ \psi^{(1)}}[-1]\,.
\label{c0v}
\eeq
\begin{figure}[t]
\centering 
\includegraphics[width=.7\textwidth,clip]{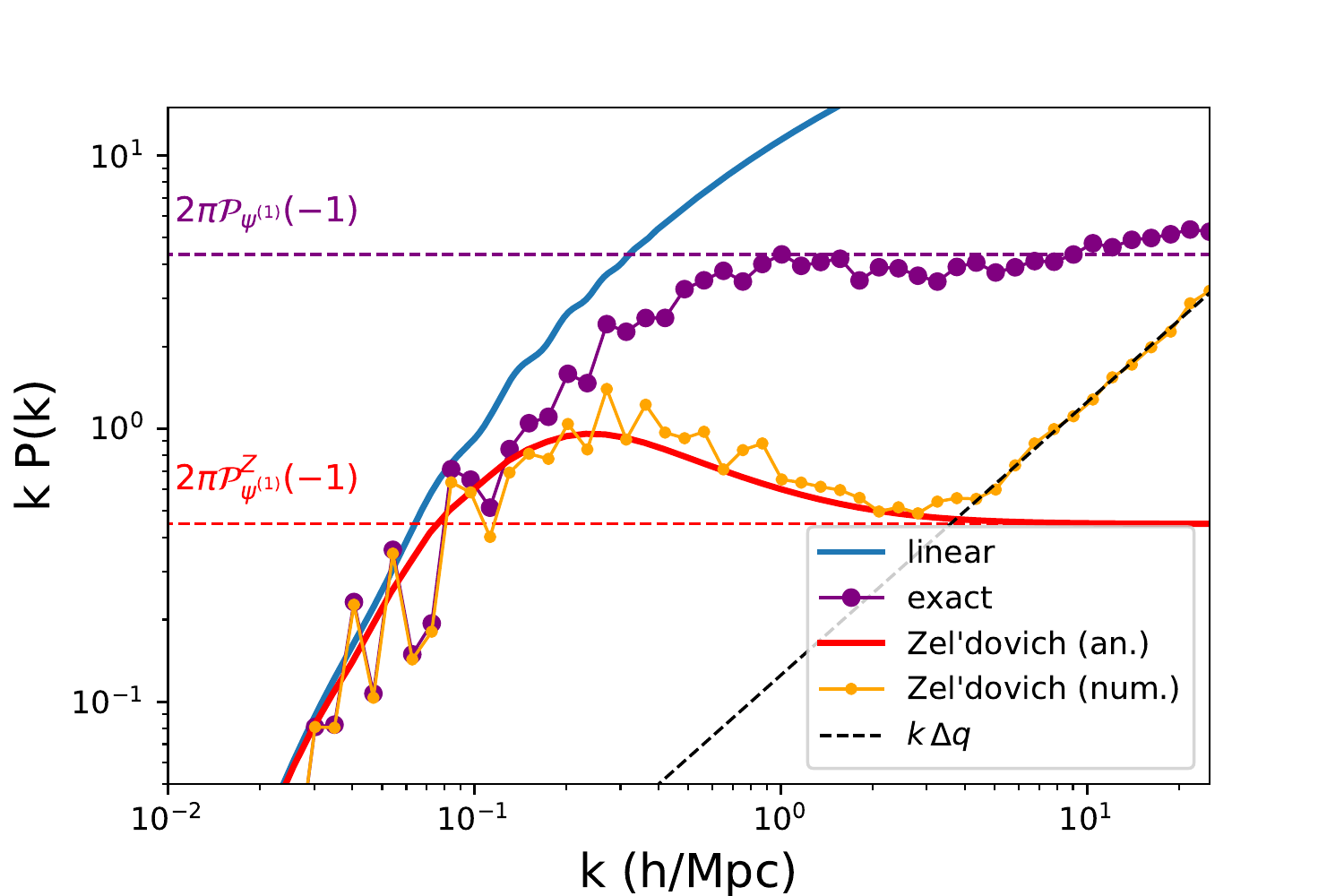}
\includegraphics[width=.7\textwidth,clip]{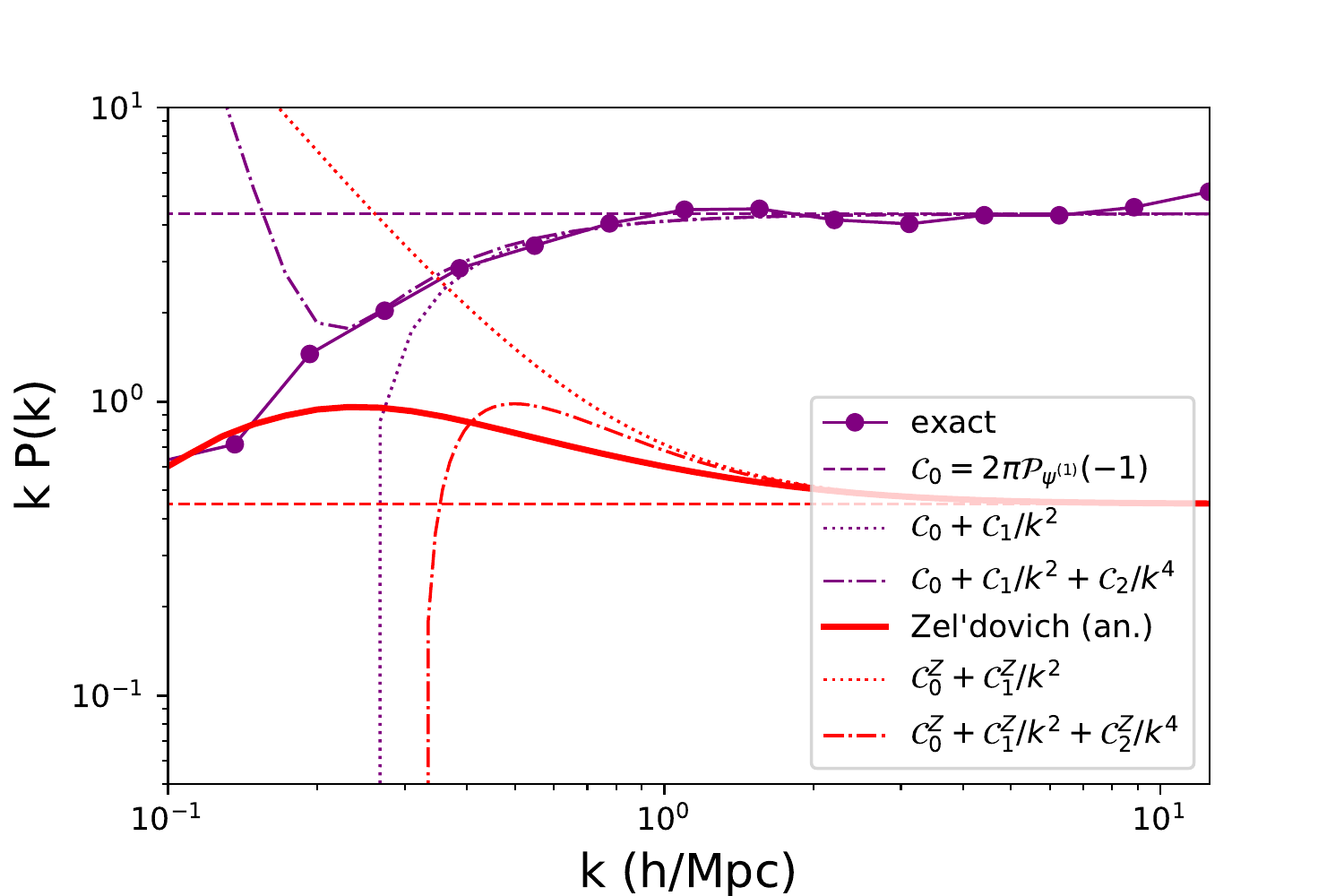}
\caption{Top: the product $k P(k)$ in one spatial dimensions in various approximations. The horizontal dashed lines are the asymptotic limits computed, according to Eq.~\re{c0v}, from the maxima of the pdf's of Fig.~\ref{Pdf1} for the full (purple) and the Zel'dovich (red) dynamics, respectively. The dashed black line indicates discreteness noise ($k \,\Delta q$ see Eq.~\re{Deltaq}). Bottom: the product $k P(k)$ for the exact (purple lines) and the Zel'dovich (red lines) dynamics. Also shown are the expansions of Eq.~\re{expas} at different orders.}
\label{kPk}
\end{figure}
The second coefficient is
\beqra
&&\!\!\!\!\!\!\!\!\!{\cal C}_1=\lim_{k\to\infty} \left ( \langle k^2 {\cal F}[\psi](k) \rangle- k^2 {\cal C}_0 \right) \nonumber\\
&& =\frac{i}{24} \langle \int_{-\infty}^{\infty} dy\,  y^3 \, e^{i y\, \left(1+\psi^{(1)} \right)} \psi^{(3)} \rangle =-\frac{2 \pi}{24}\left. \frac{d^3}{dc^3}\langle\psi^{(3)} \delta_D\left(c+\psi^{(1)} \right) \rangle\right|_{c=1} \,,\nonumber\\
&&\quad= \frac{2 \pi}{24}\left. \frac{d^3}{d \vp_1^3}\int d\vp_3 \,  {\cal P}_{ \psi^{(1)},  \psi^{(3)}}[\vp_1, \vp_3] \,\vp_3 \right|_{\vp_1=-1} \,,
\eeqra
where $ {\cal P}_{ \psi^{(1)},  \psi^{(3)}}[\vp_1,\vp_3]$ is the joint pdf for $ \psi^{(1)}$ and  $\psi^{(3)}$.
We also give the expression for the third coefficient, ${\cal C}_2 $, which has two contributions,
\beqra
&&{\cal C}_2 =\lim_{k\to\infty} \left (\langle  k^4 {\cal F}[\psi](k)\rangle - k^4 {\cal C}_0- k^2 {\cal C}_2 \right) \nonumber\\
&&=\frac{-1}{1152} \langle \int_{-\infty}^{\infty} dy\, y^6 \, e^{i y\,\left(1+ \psi^{(1)}\right) } \left(\psi^{(3)}\right)^2\rangle+\frac{i}{1920}\langle \int_{-\infty}^{\infty} dy\,  y^5 \,  e^{i y\, \left(1+ \psi^{(1)}\right)} \psi^{(5)}\rangle\,,\nonumber\\
&&\!\!\!\!\!\!  \!\!\!\!\!\! \!\!\!\!\!\! \!\!\!\!\! \!\!\!\!\!=\frac{2\pi}{1152}\left.\frac{d^6}{dc^6} \langle \left(\psi^{(3)}\right)^2 \delta_D\left(c+\psi^{(1)} \right) \rangle\right|_{c=1} +\frac{2\pi}{1920}\left.\frac{d^5}{dc^5} \langle \psi^{(5)}\delta_D\left(c+\psi^{(1)} \right) \rangle\right|_{c=1} \,,\nonumber\\
&&\!\!\!\!\!\!  \!\!\!\!\!\! \!\!\!\!\!\!  \!\!\!\!\!\!\!\!\!\! =\frac{2 \pi}{1152}\left. \frac{d^6}{d\vp_1^6}\int d\vp_3 \,  {\cal P}_{ \psi^{(1)},  \psi^{(3)}}[\vp_1, \vp_3] \,\vp_3^2 \right|_{\vp_1=-1}-\frac{2 \pi}{1920}\left. \frac{d^5}{d\vp_1^5}\int d\vp_5 \,  {\cal P}_{ \psi^{(1)},  \psi^{(5)}}[\vp_1, \vp_5] \,\vp_5 \right|_{\vp_1=-1}\,, \nonumber\\
\label{cn}
\eeqra
where the the joint pdf for $ \psi^{(1)}$ and  $\psi^{(5)}$, i.e. $ {\cal P}_{ \psi^{(1)},  \psi^{(5)}}[\vp_1, \vp_5]$, also appears.

\subsection{Asymptotic behavior in Zel'dovich dynamics}

Tthe ${\cal C}_n$  coefficients can be computed analytically in the case of Zel'dovich dynamics as the displacement field and its derivatives are gaussian. For instance, using the expression for the $\psi^{(1)}$ pdf,
\beq
 {\cal P}^{Z}_{ \psi^{(1)}}[\vp_1]= \frac{e^{-\frac{\vp_1^2}{2 \sigma_\delta^2}}}{\sqrt{2 \pi \sigma_\delta^2}}\,,
\eeq
and inserting it  in   \re{C0} gives
\beq
{\cal C}_0^{Z}= e^{-\frac{1}{2 \sigma_\delta^2}} \,\sqrt{\frac{2\pi}{\sigma_\delta^2}}   \,,
\label{c0Z}
\eeq
which is nonperturbative and non-analytic in  the variance of the linear density contrast $\sigma_\delta^2$ defined in \re{sigmas}. The correlation function counterpart of this asymptotic behavior of the PS is the expected (see discussion below Eq.~\re{expdiv}) logarithmically divergent term as $r \ll \sigma_{\Delta \psi}(\infty)/\sigma_\delta$, see Eq.~\re{logdiv},
\beq 
\!\!\!\!\ \!\!\!\!\!\!\!\!\!\!\!\!\!\!\!\!\!\xi(r) \sim 2 \int_{\frac{ \sigma_\delta}{  A \sigma_{\Delta\psi}(\infty)}}^{\infty} \frac{dk}{2 \pi}\,\cos(k r) \, P_{Z}(k) \sim \frac{e^{-\frac{1}{2 \sigma_\delta^2}}}{\sqrt{2\pi \sigma_\delta^2}} \left( \log\left(\frac{ \sigma^2_{\Delta\psi (\infty)}}{ r^2 \sigma_\delta^2}\right) +O(r^0) \right)\,,
\eeq
where we have used
\beq
P_{Z}(k) \sim \frac{e^{-\frac{1}{2 \sigma_\delta^2}}}{k} \,\sqrt{\frac{2\pi}{\sigma_\delta^2}} \,,
\eeq
and $\sigma_{\Delta \psi}(\infty)$ is defined in Eq.~\re{sigmas}.
Analogously, we can compute the next two coefficients as
\beqra
&&{\cal C}_1^{Z}= e^{-\frac{1}{2 \sigma_\delta^2}} \,\sqrt{\frac{2\pi}{\sigma_\delta^2}} \;\frac{1-6\sigma_\delta^2+3 \sigma_\delta^4}{24\, \sigma_\delta^8}\, \sigma_{13}\,,\label{c1Z}\\
&&{\cal C}_2^{Z}= e^{-\frac{1}{2 \sigma_\delta^2}} \,\sqrt{\frac{2\pi}{\sigma_\delta^2}} \;\Big[ \frac{1-28\sigma_\delta^2+210\sigma_\delta^4-420\sigma_\delta^6 +105\sigma_\delta^8}{1152\,\sigma_\delta^{16}} \,\sigma_{13}^2\nonumber\\
&& \qquad \qquad\qquad\qquad+\frac{1-15\sigma_\delta^2 +45\sigma_\delta^4-15\sigma_\delta^6}{720\,\sigma_\delta^{12}}\,\sigma_{33}\Big]\,,\label{c2Z}
\eeqra
where
\beqra
&&\sigma_{13}\equiv -\langle \psi^{(1)}(0) \psi^{(3)}(0)\rangle = \int\frac{dp}{2\pi} P_{\rm{lin}}(p) p^2\,,\nonumber\\
&&\sigma_{33}\equiv \langle \left( \psi^{(3)}(0)\right)^2\rangle = \langle \psi^{(1)}(0) \psi^{(5)}(0)\rangle =  \int\frac{dp}{2\pi} P_{\rm{lin}}(p) p^4\,.
\eeqra
Notice that the integrals defining $\sigma_\delta^2$, $\sigma_{13}$ and $\sigma_{33}$ are all UV divergent, requiring a cutoff be introduced. The dependence of our results on this smoothing scale is discussed below in section \ref{smoothdisc}.

\subsection{Asymptotic Behavior in Full dynamics}
We now analyze the asymptotic behavior in the case of full dynamics by solving the equation of motion, Eq.~ \re{fulleq}, with initial conditions \re{vin}, \re{incond}, and \re{incond2} imposed at some early time  on the displacement field $\psi(q,\tau_{\rm{in}})$ and its time derivative. To get the initial PS, we take a $\Lambda$CDM PS at $z=0$ obtained by CAMB  \cite{CAMB},  define a corresponding 1D PS as \cite{McQuinn:2015tva}
\beq
P_{\rm{linear}}(k)= \frac{k^2}{2\pi}P_{\rm{CAMB}}(k)\,,
\eeq
and rescale it at the initial redshift assuming Einstein de Sitter cosmology, that is, dividing by $(1+z_{\rm{in}})^2$  (we will set $z_{\rm{in}}=99$).
We solve the equations on a periodic line of size L=6000 Mpc/h, discretized on a grid of  $2N=48,000$ points and with 500 time steps, using the algorithm described in \cite{Pietroni:2018ebj} and in \ref{numerics}. As a check, we solve also for the Zel'dovich dynamics, using the force \re{ZeldF}, with the same initial conditions imposed, and compare with the analytic results presented in the previous subsection. 

\begin{figure}[t]
\centering 
\includegraphics[width=.45\textwidth,clip]{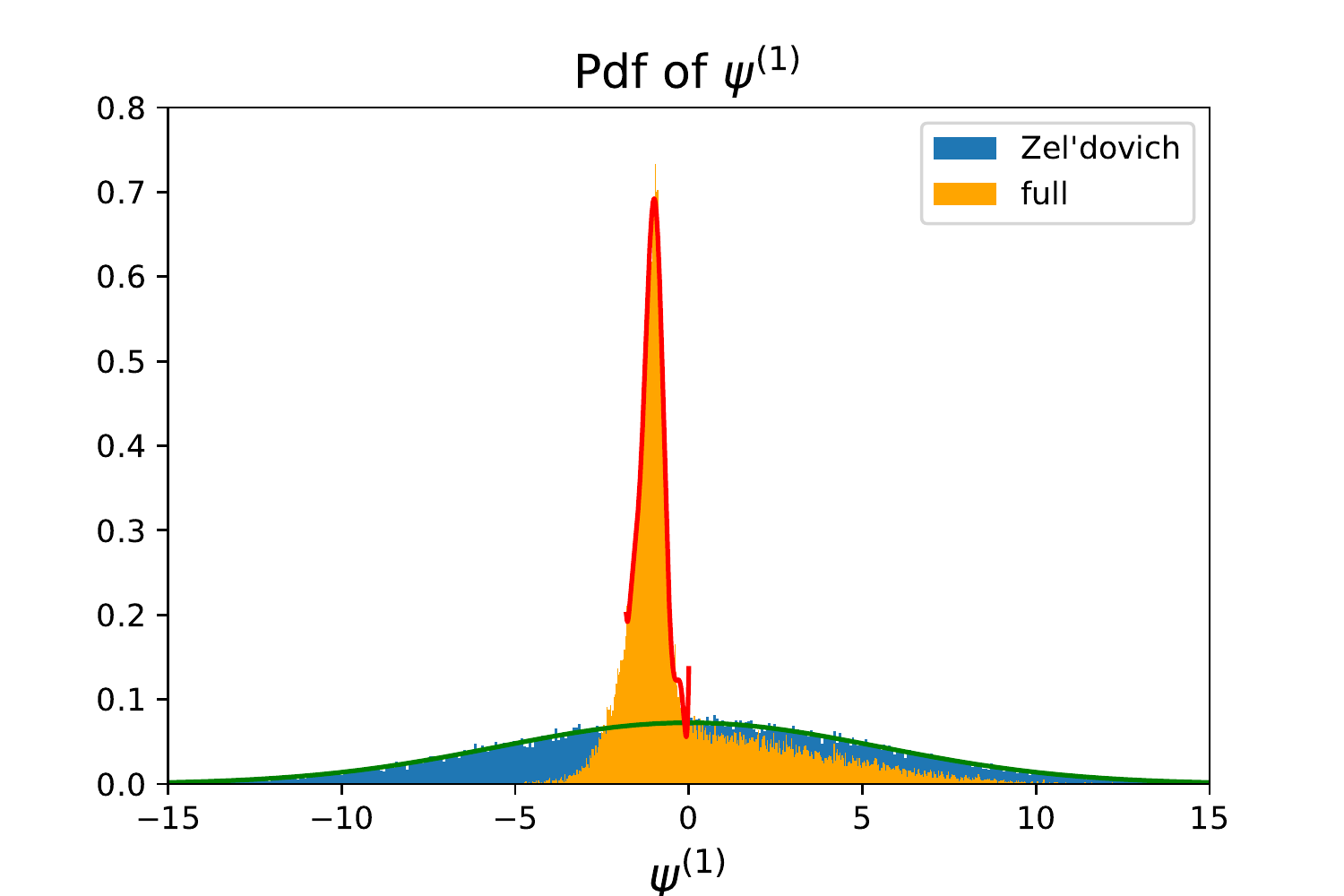}
\includegraphics[width=.45\textwidth,clip]{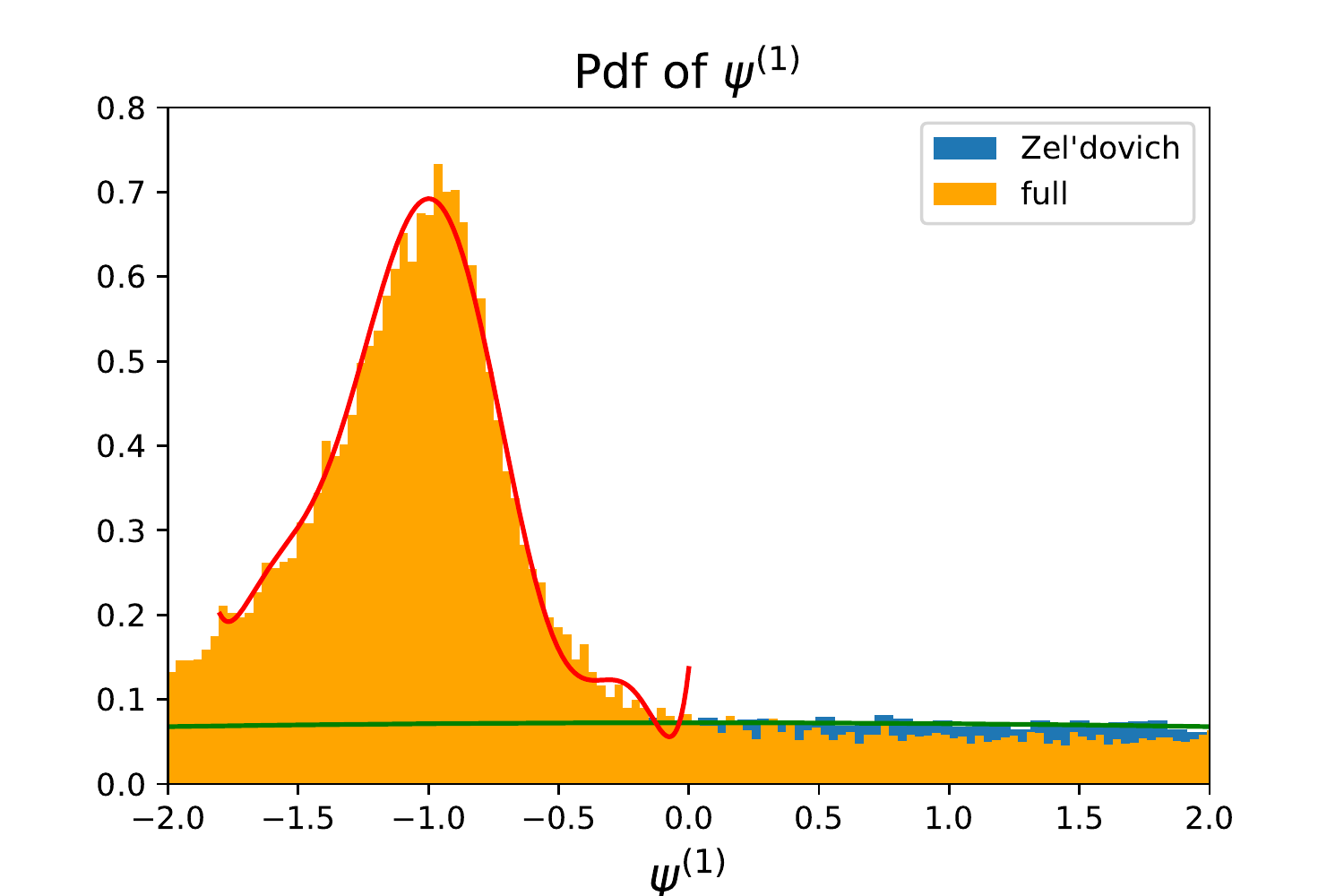}
\caption{Left: the Pdf for $\psi^{(1)}$ for the real dynamics (orange) and the Zel'dovich one (blue). The green curve is the analytic, gaussian, result for the Zel'dovich dynamics, while the red curve is a polynomial fit around the maximum of the histogram obtained for the full dynamics. Right: a zoom of the left panel. Notice that the maximum for the real dynamics is at $\psi^{(1)}\simeq -1$, as expected from the attractor behavior discussed in the text.  }
\label{Pdf1}
\end{figure}

\begin{figure}[t]
\centering 
\includegraphics[width=.45\textwidth,clip]{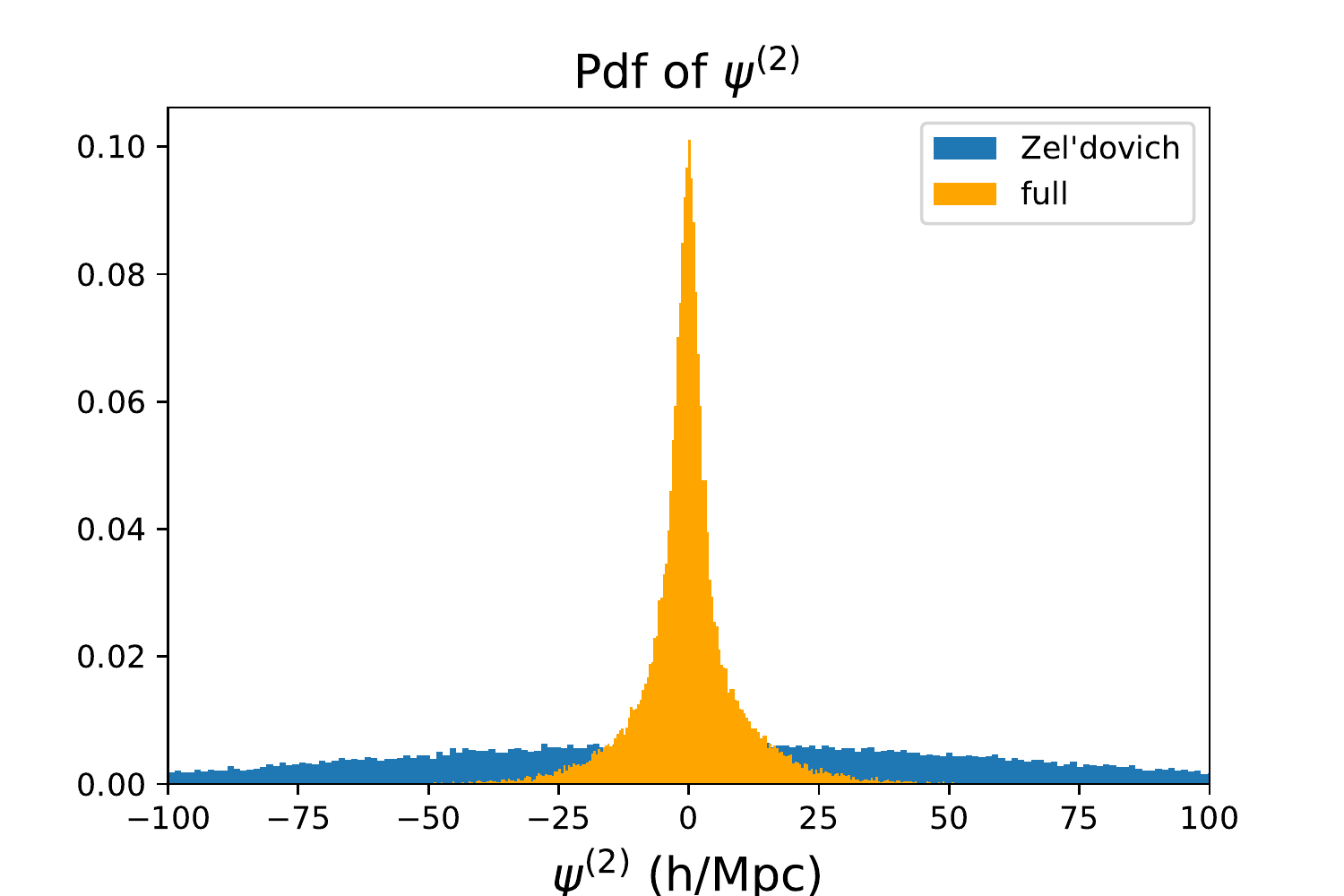}
\includegraphics[width=.45\textwidth,clip]{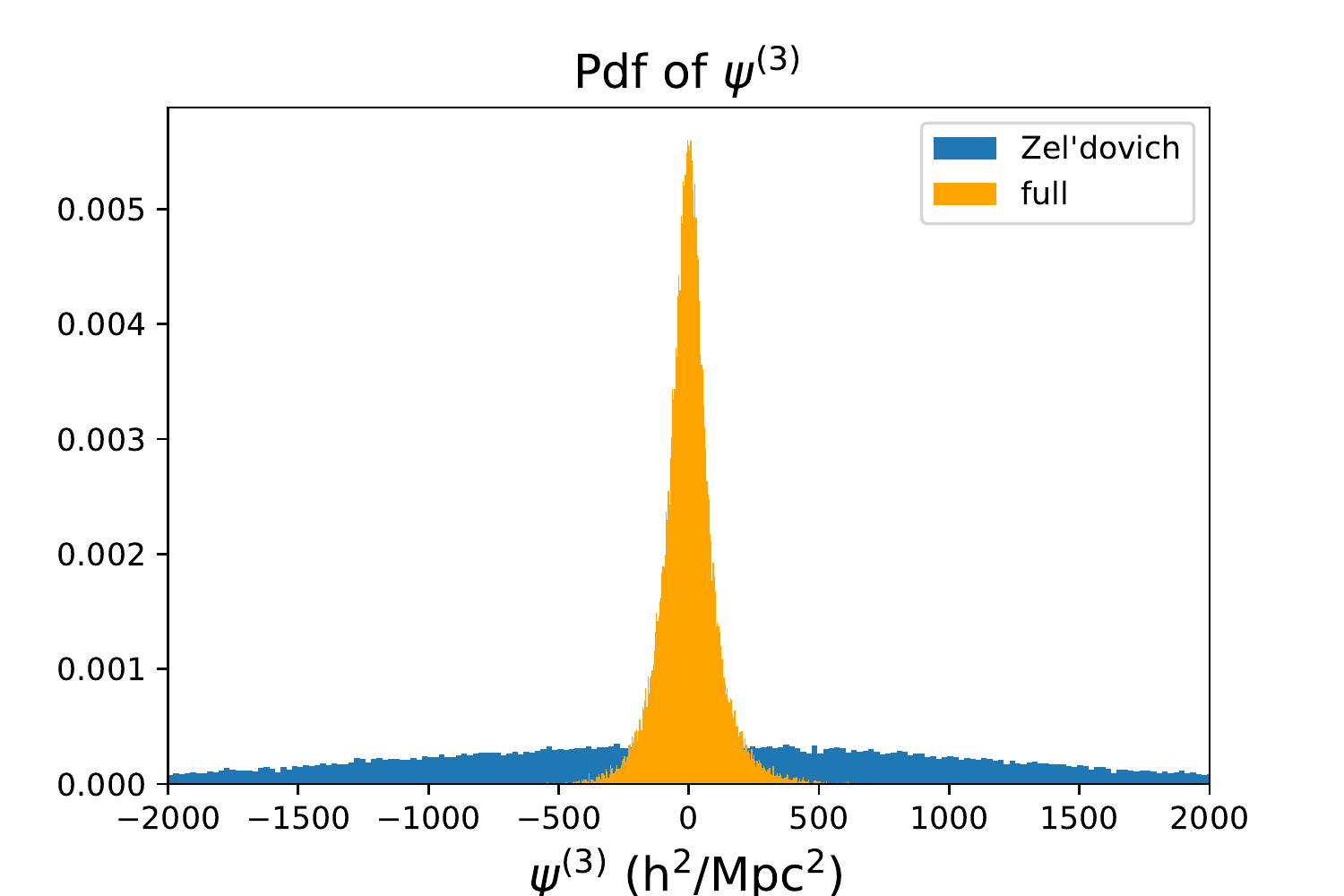}
\caption{Left: the Pdf for $\psi^{(2)}$ for the real dynamics (orange). Right: the Pdf for $\psi^{(3)}$ for the real dynamics (orange)  }
\label{Pdf23}
\end{figure}

In Fig.~\ref{numsol} we show, in the left panel, the mapping between Lagrangian and Eulerian space, that is, the function
\beq
x(q)=q+\psi(q)\,,
\eeq
evaluated at $z=0$, and, in the right panel, the Eulerian phase-space diagram, that is $\dot\psi(q)/{\cal H} =v(q)/{\cal H}$ vs $x(q)$. The left panel shows clearly the most prominent feature of full dynamics, compared to Zel'dovich one, namely, the flattening of the mapping inside multistreaming regions. As shown analytically in \cite{Pietroni:2018ebj},  inside multistreaming regions the full dynamics exhibits attractor behavior, such that the first and all higher order derivatives of $x(q)$ with respect to $q$ tend to vanish, where the value of $x(q)$ itself follows the center of mass of the matter inside the  region. The attractor is nicely confirmed by the inspection of the pdf's for $\psi^{(1)}(q)$, $\psi^{(2)}(q)$, and, $\psi^{(3)}(q)$, shown in Figs.~\ref{Pdf1} and \ref{Pdf23}. In particular, while for the Zel'dovich dynamics these derivatives follow the expected gaussian distributions, the distribution of $\psi^{(1)}(q)$ for the full dynamics is clearly non-gaussian and peaked at $\psi^{(1)}(q) \simeq-1$, that is, at $x^{(1)}(q)=0$. Interestingly, $\psi^{(2)}(q)$ and $\psi^{(3)}(q)$ can still be fit by a gaussian centered around 0, though with a much smaller variance than the Zel'dovich one.

From \re{C0} we know that the value of the pdf for $\psi^{(1)}(q)$ in $-1$ is directly related to the coefficient of the leading term in the large $k$ expansion, which now can be predicted directly from Fig.~\ref{Pdf1}.  This is shown in Fig.~\ref{kPk}, where we plot, on the upper panel, the product $k P(k)$ in linear theory (blue), in the exact dynamics (purple) and in the Zel'dovich one (red for the analytic result, orange for the numerical one). The horizontal dashed lines show the expected asymptotic behaviors. For the full dynamics, the limiting value is obtained by reading $ {\cal P}_{ \psi^{(1)}}[-1] $ from a polynomial fit to the histogram in Fig.~\ref{Pdf1} around the peak, while for the Zel'dovich dynamics it is computed via the analytical result of Eqs.~\re{c0v}, \re{c0Z}, checked to be consistent with the numerical one obtained from the Zel'dovich histogram in Fig.~\ref{Pdf1}.

In Fig.~\ref{kPk} we also show, with the black-dashed line, the expected value of the  noise due to the discreteness of the spatial grid,
\beq
k P_{\rm{d}}(k)=k \,\Delta q= k \frac{L}{2 N}\,,
\label{Deltaq}
\eeq
where $2 N$ is the number of grid points. This can be understood by discretizing Eq.~\re{funcF},
\beq
{\cal F}[\Delta \psi](k) \to k \,\Delta q \sum_{n=-N}^{N-1}  e^{i k\left(n \Delta q + \Delta\psi(n \Delta q)\right) }\,,
\eeq
and considering the  large $k$ regime, where  we get,
\beqra
&&\!\!\!\!\!\!\!\! \!\!\!\!\!\!\!\! \!\!\!\!\!\!\!\!   \!\!\!\!\!\!\!\!   k \,\Delta q \sum_{n=-N}^{N-1}  e^{i k n \Delta q\left(1  + \psi^{(1)}(0)\right) }= k \,\Delta q \cot\left(\frac{k \Delta q \left(1  + \psi^{(1)}(0)\right)}{2}\right) \sin\left(\frac{k L \left(1  + \psi^{(1)}(0)\right)}{2}\right)\nonumber\\
&&\!\!\!\!\!\!\!\! \!\!\!\!\!\!\!\!  \!\!\!\!\!\!\!\! \to \pi \delta_D\left(\frac{1+\psi^{(1)}(0)}{2}\right)\qquad\qquad (\Delta q\to 0\,,\quad L\to \infty)\,. 
\label{disc}
\eeqra

\begin{figure}[t]
\centering 
\includegraphics[width=.55\textwidth,clip]{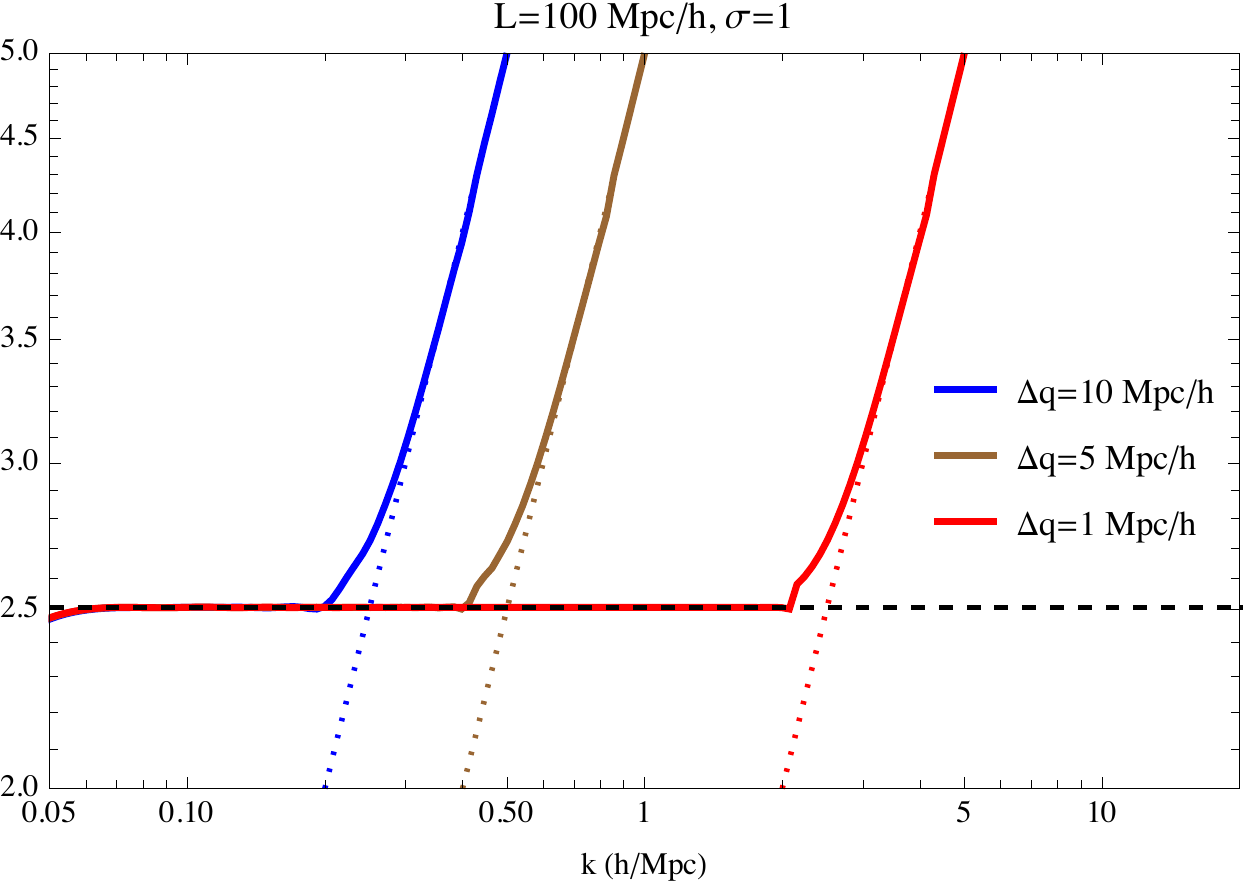}
\caption{The effect of a finite grid size $\Delta q \neq 0$ on the PS. To make this plot, we have used a gaussian pdf centered in $\psi^{(1)}(0)=-1$, with unitary standard deviation and the discretized form for the ${\cal F}[\Delta\psi](k)$ functional of Eq.~\re{disc}. The solid lines give the result of the numerical integrals in  $\psi^{(1)}(0)$ for different values of $\Delta q$, the dotted lines are the corresponding $k \Delta q$ contributions, while the horizontal dashed line gives $2 \pi {\cal P}_{\psi^{(1)}}(-1)$. The linear ``volume" has been set to $L=100$ Mpc/h. }
\label{PP3}
\end{figure}

The effect of a finite $\Delta q$ is shown in Fig.~\ref{PP3}, where we plot the results of a numerical integration in $\psi^{(1)}(0)$, using the first line of Eq.~\re{disc} and a gaussian pdf centered in $\psi^{(1)}(0)=-1$, with unitary standard deviation. As we see, both the expected plateau, $2 \pi {\cal P}_{\psi^{(1)}}(-1)$, and the noise contributions $k \Delta q$ are correctly reproduced and the transition between the two behaviors agree with that of the numerical results in the Zel'dovich dynamics (orange line in Fig.~\ref{kPk}). Notice that the asymptotic value of the noise, Eq.~\re{Deltaq}, comes from the $n=0$ term in the sum at the LHS of Eq.~\re{disc}.

The large-$k$ plateau  is very clear for the exact dynamics  in Fig.~\ref{kPk}, thanks to the fact that the peak of the pdf in $-1$ raises it well above the discreteness noise contribution, which is not the case for the (numerical) Zel'dovich result. The pdf peak dominates the PS behavior down to quite small values of $k$, whereas in the Zel'dovich dynamics, for which the pdf is not peaked in -1, the contributions of higher order contributions in $1/k$ push the onset of the asymptotic regime to  larger $k$'s. In the bottom panel of Fig.~\ref{kPk} we plot $k P(k)$ along with different terms in the expansion \re{expas} for the PS. For the Zel'dovich dynamics the coefficients of the expansion are computed from  Eqs.~\re{c0Z}, \re{c1Z}, and \re{c2Z}, while for the full dynamics they are obtained by a fit to the PS.

To estimate the range of validity of the $1/k$ expansion we require that higher order terms are subdominant with respect to lower order ones, which gives the criteria
\beq
k^2> \frac{{\cal C}_1}{{\cal C}_0}, \; \frac{{\cal C}_2}{{\cal C}_1}\,\cdots. 
\eeq
For Zel'dovich dynamics, these criteria typically imply that for  $k < 2 \pi \sigma_\delta/\sigma_{\Delta\psi}(\infty)$ (which turns out to be $ \simeq 2.4 \,\rm{h/Mpc}$ for the parameters chosen in our simulation), all the terms of the expansion are of the same order. Judging from Fig.~\ref{kPk}, the expansion for the full dynamics can  be  extended to somewhat smaller values of $k$, but  not as low as to match the SPT range,  $k\alt 0.1 \, \rm{h/Mpc}$.

\subsection{Dependence on the Lagrangian smoothing scale}
\label{smoothdisc}
\begin{figure}[t]
\centering 
\includegraphics[width=.45\textwidth,clip]{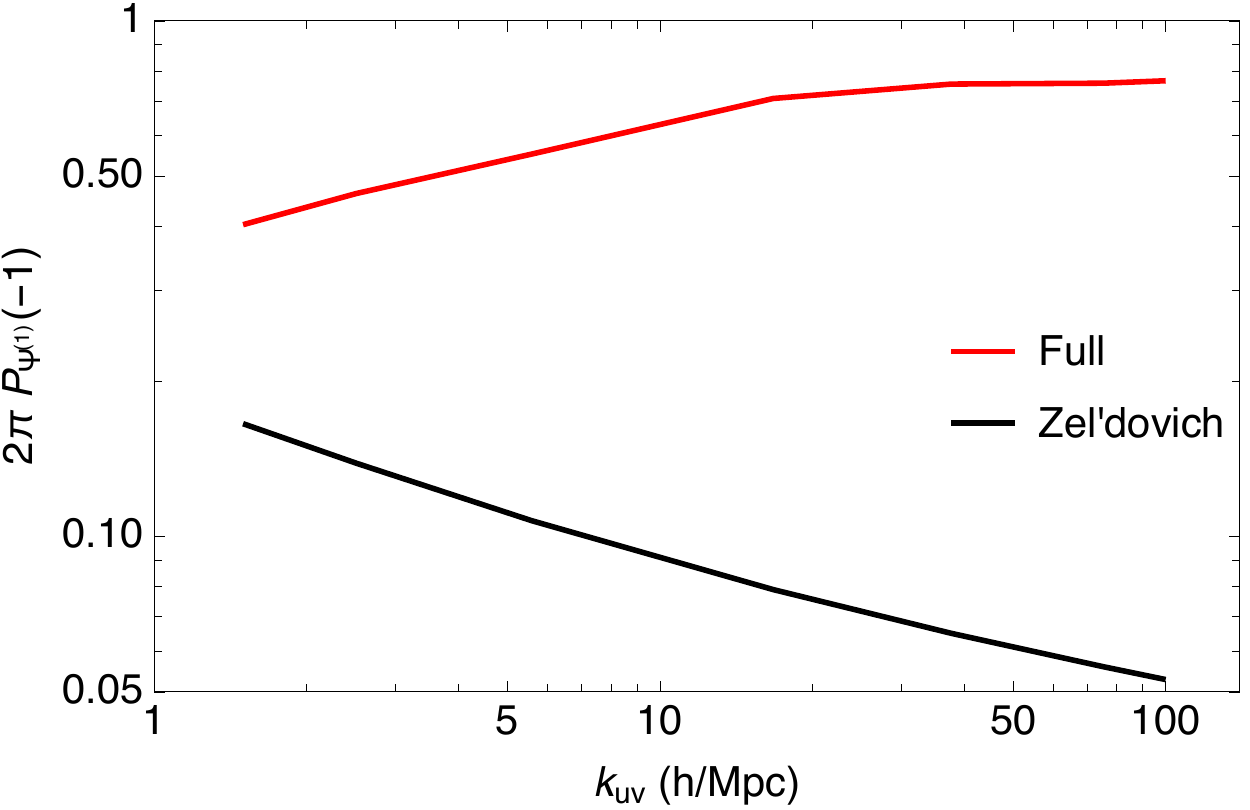}
\caption{Dependence of $2\pi\, {\cal P}_{ \psi^{(1)}}[-1]$  on the smoothing scale, $k_{\rm{uv}}$, for the full dynamics and for the Zel'dovich one.  }
\label{smoothdep}
\end{figure}

In any practical setting, the definition of the displacement field implies a smoothing scale. It can be identified with the spatial resolution in Lagrangian space, or with the inverse of the maximum Fourier mode in \re{Fpsi}. In Zel'dovich dynamics the coefficient of the $1/k$ term in Eq.\re{c0Z} depends on the smoothing length through the dependence of 
$\sigma_\delta^2$ on the momentum cutoff of the integral defining it, i.e. Eq.~\re{sigmas}. Indeed, as is well known, the integral is UV-divergent for a $\Lambda$CDM linear PS, and in order to give a meaningful definition to the variance of $\delta$ one should introduce a UV regulator, like, for instance, a gaussian one,
\beq
 \sigma_\delta^2(k_{\rm{uv}}) =\int\frac{dp}{2\pi}\, P_{\rm{lin}}(p) e^{-\frac{p^2}{k_{\rm{uv}}^2}}\,.
 \label{varZ}
\eeq
At $z=0$, one has $ \sigma_\delta^2(k_{\rm{uv}})\gg 1$ for $k_{\rm{uv}}\gg1$, such that
\beq
{\cal C}_0^{Z}= e^{-\frac{1}{2  \sigma_\delta^2(k_{\rm{uv}})}} \,\sqrt{\frac{2\pi}{ \sigma_\delta^2(k_{\rm{uv}})}}\simeq \sqrt{\frac{2\pi}{ \sigma_\delta^2(k_{\rm{uv}})}}\,,
\eeq
i.e. the gaussian distribution for $ \psi^{(1)}$ becomes wider and wider  for increasing $k_{\rm{uv}}$, and, as a consequence, its value at  $\psi^{(1)}=-1$ decreases. 
In the case of full dynamics the smoothing scale dependence can be studied only numerically. We identify $k_{\rm{uv}}= N \pi/L$ (where $N$ is the number of grid points), as this correspondence provides the correct matching between the analytical result \re{varZ} and the numerical one in the case of the Zel'dovich dynamics. 
In Fig.~\ref{smoothdep} we show the dependence on $k_{\rm{uv}}$ of the quantity $2\pi\, {\cal P}_{ \psi^{(1)}}[-1]$ (that is, of ${\cal C}_0$) for full and Zel'dovich dynamics. While in the Zel'dovich case the dependence on the smoothing scale persists down to the smallest scales available, in the exact case it reaches a plateau, which can be interpreted as another manifestation of the post shell-crossing attractor.

\section{From 1+1 to 3+1}
\label{3d}
The  $3+1$ dimensional version of the expression \re{psd} for the PS is,
\beq
P(k)+ (2\pi)^3\delta_D(\bk)=\langle \int d^3q\, e^{i \bk\cdot\left( \bq+{\bf{\Delta \Psi} (\bq)}\right)}\rangle \,,
\label{P3D}
\eeq
where now the relative displacement vector field is given by
\beq
{\bf{\Delta \Psi}}(\bq)= {\bf \Psi}\left(\frac{\bq}{2}\right) -{\bf \Psi}\left(-\frac{\bq}{2}\right) \,.
\eeq
At large $k$ we can again consider a small $q$ expansion,
\beq
\bk\cdot\left( \bq+{\bf{\Delta \Psi} (\bq)}\right)\simeq k^i q_j\left(\delta^i_j+M^i_j \right) +\frac{1}{24}k_i q^jq^kq^l B^i_{jkl} +\cdots\,,
\eeq
where the first two terms are given by
\beq
M^i_j\equiv \frac{\partial \Psi^i(0)}{\partial q^j}, \quad B^i_{jkl}\equiv \frac{\partial^3 \Psi^i(0)}{\partial q^j\partial q^k \partial q^l},
\eeq
with $M$ the deformation tensor.
By defining ${\bf y} = k\,\bq$, the PS at large $k$ can be approximated as
\beq
P(k)\simeq \frac{1}{k^3} \langle \int d^3y\, e^{i \hat k_i y^j\left(\delta^i_j+M^i_j\right)+ \frac{i}{24} \hat k_iy^jy^ky^l\,\frac{B^i_{jkl}}{k^2}+\cdots} \rangle\,,
\label{exp3d}
\eeq
where $\hat k\cdot\hat k=1$.
The leading term in the large-$k$ limit is then
\beq
P(k)\sim \frac{1}{k^3} \langle \int d^3 y \,e^{i \by\cdot {\bf V}}\rangle = \frac{(2 \pi)^3}{k^3} \langle \delta_D\left({\bf V}\right)\rangle\,,
\eeq
where 
\beq
V_j\equiv   \hat k_i \left(\delta^i_j+M^i_j\right)\,.
\eeq
Setting  $\hat k$ along the $z-$axis without loss of generality we can express the expectation value in terms of the joint pdf of the three components of the deformation tensor, $M^3_j$, namely,
\beqra
&&\!\!\!\!\!\!\!\!\! P(k)\sim \frac{(2 \pi)^3}{k^3} \int dx_1 dx_2 dx_3 {\cal P}_{M^3_1,M^3_2,M^3_3}[x_1,x_2,x_3] \delta_D(x_1) \delta_D(x_2) \delta_D(1+x_3),\nonumber\\
&&\quad  = \frac{(2 \pi)^3}{k^3} {\cal P}_{M^3_1,M^3_2,M^3_3}[0,0,-1] \,.
\label{asi3}
\eeqra
The result above has a nice physical interpretation: the coefficient of the leading term of the asymptotic expansion is  given by the value assumed by the pdf on configurations that exhibit shell crossing on a plane, that is, on {\it pancakes} configurations.

\begin{figure}[t]
\centering 
\includegraphics[width=.55\textwidth,clip]{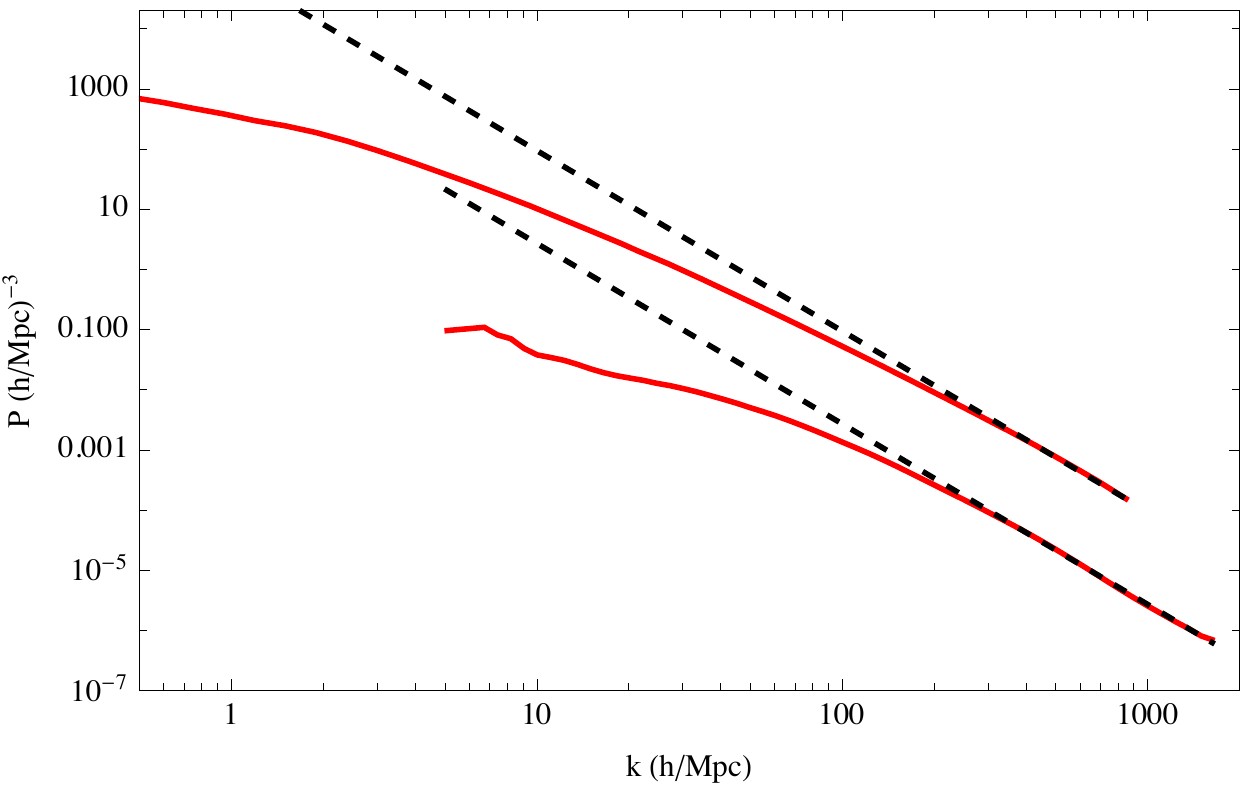}
\includegraphics[width=.55\textwidth,clip]{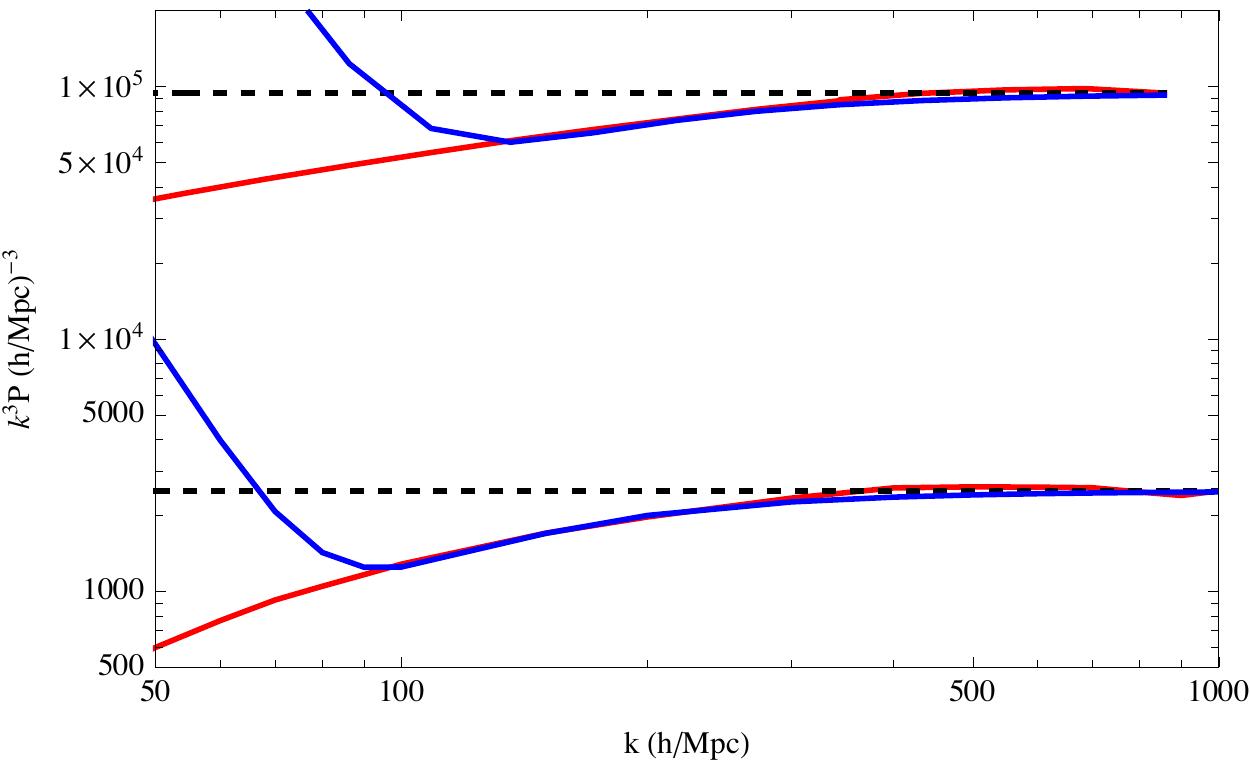}
\caption{Top: The CDM PS in 3 spatial dimensions at very large $k$'s. Data are taken from the simulations of \cite{Mocz:2019uyd,  Mocz:2019emo} at $z=7$ (lower red line) and from the  Illustris-TNG-100 simulation \cite{Springel:2017tpz} (upper red line) at $z=0$. The black dashed lines scale as $k^{-3}$, as predicted by the asymptotic solution, Eq.~\re{asi3}. Bottom: The same data, in which the PS's have been multiplied by $k^3$. The blue lines are obtained by adding the first two next-to-leading terms, $1/k^5$ and $1/k^7$,  with a fitted coefficient.}
\label{Plot3D}
\end{figure}

The $1/k^3$ asymptotic behavior predicted by Eq.~\re{asi3} seems to be confirmed in data from high resolution simulation. In Fig.~\ref{Plot3D} we show, on the top panel, the results from Mocz et al, \cite{Mocz:2019uyd,Mocz:2019emo} on the CDM PS at $z=7$, and from the CDM-only simulation from the Illustris-TNG-100 suite at $z=0$ \cite{Springel:2017tpz}. In both cases, the $1/k^3$ behavior is attained for $k>O(500)\,{\rm h/Mpc}$. On the bottom panel we plot $k^3 P(k)$ for the same data, and, with blue lines, the asymptotic expansion including the $1/k^3$ and $1/k^5$ terms, with fitted coefficients.

On the analytical side, we can also compute the asymptotic coefficients within the  Zel'dovich approximation. The deformation tensor in this case is given by
\beq
M^i_{j,Z}= \frac{1}{{\cal H} f}\frac{\partial v^i(0)}{\partial q^j}\,,
\eeq
where $v^i$ is the velocity in linear perturbation theory. It is then straightforward to compute the correlators entering the pdf, 
\beq
\langle M^i_{j,Z} M^l_{k,Z} \rangle= \frac{1}{15} \sigma_\delta^2 \left(\delta^i_j \delta^l_k +\delta^i_k \delta^l_j +\delta^{il}\delta_{jk} \right)\,,
\label{Mzd}
\eeq
where now
\beq
\sigma_\delta^2\equiv \int \frac{d^3 k}{(2 \pi)^3} P(k)\,e^{-\frac{k^2}{k_{\rm uv}^2}}.
\eeq
The pdf is then given by
\beq
{\cal P}^{Z}_{M^3_1,M^3_2,M^3_3}[x_1,x_2,x_3] =\frac{\left(\det \Sigma_M\right)^{-1/2}}{(2 \pi)^{3/2}} e^{-\frac{1}{2} X^T\cdot \Sigma_M^{-1}\cdot X}\,,
\eeq
where
\beq
\Sigma_M =\frac{\sigma_\delta^2}{15} \left( 
\begin{tabular}{ccc}
1&0&0\\
0&1&0\\
0&0&3
\end{tabular}
\right)\,,
\qquad X= \left(
\begin{tabular}{c}
$x_1$\\
$x_2$\\
$x_3$
\end{tabular}
\right)\,.
\eeq
Using this expression in \re{asi3} gives
\beq
P^{Z}(k)\sim \frac{15\sqrt{5}}{k^3} \left(\frac{2 \pi}{\sigma_\delta^2}\right)^{3/2} e^{-\frac{5}{2 \sigma_\delta^2}}\,.
\eeq
The same result can be obtained by inverting the average and the integral in \re{exp3d} (with $B^i_{jkl}=0$), taking the gaussian average of the exponential, which gives
\beq
P^{Z}(k)\sim \frac{1}{k^3} \int d^3y \,e^{i \hat k_i y^i} e^{-\frac{1}{2} \hat k_i \hat k_l y^j y^k \langle M^i_j M^l_k\rangle }\,,
\eeq
and using  Eq.~\re{Mzd} to compute the correlator in the exponent.

\section{Link to the halo model}
\label{halolink}
In this final section, we want to investigate the relation between the asymptotic behavior of the PS discussed in the previous sections and the halo model (for a review, see \cite{Cooray:2002dia}), which is used extensively to model power spectra beyond the perturbative regime. The basic hypothesis of the halo model is that CDM is organized in halos of different masses, in the sense that any CDM particle belongs to an unambigously defined  halo of a given total mass $M$. The halos are collectively described by a mass function $n(M)$, normalized such that
\beq
\int dM\;M\,n(M)=\bar \rho\,,
\eeq
where $\bar \rho$ is the mean CDM density. Moreover, the halo model  (at least in its simplest version) assumes that the density profile of a halo depends only on the halo mass, 
\beq
\rho(\bx)=M u(\bx|M)\,,
\eeq
where $\bx$ is the position relative to the halo center, and $u(\bx|M)$ is the halo profile function, normalized such that $\int d^3x\, u(\bx|M)=1$. We will also make the extra simplifying assumption that halos are spherical. 

Given the above hypotheses, the correlation function and the PS can  be split in two parts, the ``1-halo'' and the ``2-halo'' terms, the former due to contributions from pairs of particles belonging to the same halo and the latter from particles belonging to different halos.  At short scales, the 1-halo term dominates, and will therefore be the relevant contribution in the $k\to \infty$ limit we are considering.
The 1-halo term is given by
\beq
\xi_{\rm 1h}(r)=\frac{1}{\bar\rho^2} \int dM\;M^2\,n(M)\int d^3y\;u(\by-\br/2)|M)u(\by+\br/2)|M)
\eeq
and
\beq
P_{\rm 1h}(k)=\frac{1}{\bar\rho^2} \int dM\;M^2\,n(M) \left| \tilde u(\bk|M)\right|^2\
\label{P1hst}
\eeq
for the correlation function and PS, respectively, where $ \tilde u(\bk|M)$ is the Fourier transform of $u(\bx|M)$.

On the other hand, an analogous splitting can be performed starting from Eq.~\re{P3D}. The halo model hypotheses imply that the pdf of the displacement field can be represented as the integral
\beq
{\cal P}[{\bf \Psi}]= \int dM\;\frac{M^2}{\bar\rho^2}\,n(M)\;{\cal P}[{\bf \Psi} |M]\,,
 \label{normP}
\eeq
where ${\cal P}[{\bf \Psi} |M] = {\cal P}[{\bf \Psi}]$ if, on the field configuration ${\bf \Psi}(\bq)$, the origin $\bq=0$ is  contained in a halo of mass $M$ (and profile $u(\bx|M)$)  while ${\cal P}[{\bf \Psi} |M] =0$ otherwise.
In this case, we can give a path integral representation of  the ``1-halo'' contribution to the PS as
\beqra
&& \!\!\!\!\!\!  \!\!\!\!\!\!  \!\!\!\! \hat P_{\rm 1h}(k)=\langle  \int d^3q\, e^{i \bk\cdot\left( \bq+{\bf{\Delta \Psi} (\bq)}\right)} \rangle|_{\rm 1h}\,, \nonumber\\
&&= \int {\cal D} {\bf \Psi}  \int dM\;\frac{M^2}{\bar\rho^2}\,n(M)\;{\cal P}[{\bf \Psi} |M]\,\int d^3q\, F_{\rm 1h}(q|M)\, e^{i \bk\cdot\left( \bq+{\bf{\Delta \Psi} (\bq)}\right)}\,.
\label{P1hus}
\eeqra
 The function $F_{\rm 1h}(q|M)$ gives the probability that both the points at $\bq/2$ and at $-\bq/2$ are contained in the same spherical halo of mass $M$ and Lagrangian radius 
\beq
q_M = \left(\frac{3}{4 \pi}\frac{M}{\bar \rho}\right)^{1/3}\,,
\eeq
and it reads \cite{Valageas:2010yw}
\beq
F_{\rm 1h}(q|M)=\frac{(2 \,q_M-q)^2(4 \,q_M+q))}{16 \,q_M^3}\qquad\qquad ({\rm{for}}\;\;\;0\le q\le 2 q_M)\,,
\eeq
and zero otherwise.
By comparing \re{P1hst} and \re{P1hus} we see that the two expression coincide if we identify
\beq
 \left| \tilde u(\bk|M)\right|^2 \leftrightarrow \int {\cal D} {\bf \Psi} \;{\cal P}[{\bf \Psi} |M]\,\int d^3q\, F_{\rm 1h}(q|M) \,e^{i \bk\cdot\left( \bq+{\bf{\Delta \Psi} (\bq)}\right)}\,,
\eeq
wich relates the Eulerian space halo profiles with ensemble averages of quantities defined in Lagrangian space. 
 However, the above identification cannot hold at the very large $k$ values we are considering here. Indeed, 
proceeding as in the previous section, and noticing that $ F_{\rm 1h}(q|M) \to 1$ as $q\to 0$, we get that in the $k\to \infty$ limit, the term on the right gives
\beq
\frac{(2 \pi)^3}{k^3} {\cal P}_{M^3_1,M^3_2,M^3_3}[0,0,-1|M]\,,
\eeq
which, after integration over $M$ as in Eq.~\re{normP}, gives  Eq.~\re{asi3}. The relevant field configurations are  two-dimensional pancakes of mass $M$, quite different from the spherically symmetric halos assumed in the halo model. In other terms, the relevant configurations are absent from the pdf, once it is represented as in \re{normP}. Therefore, the simplest halo model assumption, namely, that matter is organised in spherical halos, must be generalized in order to reproduce the PS at very large $k$'s.

\section{Conclusions}
\label{fine}
In this paper we have explored the deep UV limit of the LSS PS (and correlation function) produced by CDM. The results can be summarized as follows: the PS at large $k$'s can be expressed as an expansion in  powers of the form $1/k^{d+2n}$, where $d$ is the number of spatial dimensions, and $n$ is a non negative integer. The coefficients of this expansion are related to the properties of the pdf of the displacement field around specific configurations in Lagrangian space. For instance, in $d=1$, the coefficient of the leading $1/k$ term is related to the value of the pdf computed in $\psi^{(1)}=-1$, that is, to the abundance of Lagrangian points experiencing shell-crossing. In $d=3$ the relevant configurations are those in which shell-crossing is taking place along one dimension, namely, pancakes. In principle, the coefficients depend on the smoothing scale. However, at least in $d=1$ and for a $\Lambda$CDM like linear PS, we have found that the smoothing scale dependence disappears for sufficiently fine smoothing, this behavior being related to the existence of attractors in the post shell-crossing dynamics. 

While in $d=1$ the leading asymptotic behavior sets in at $k>k_{\rm min} =O(1) \,{\rm h/Mpc}$, in the physically relevant $d=3$ case we have $k_{\rm min} =O(500) \,{\rm h/Mpc}$ , which is far away from the region of observational interest, and where, moreover, other physical  (baryon feedback, free-streaming, bias ...) and observational (shot noise...) effects would certainly play a role and have to be taken into account. Adding next-to-leading terms, like the  $1/k^5$ and $1/k^7$ contributions,  reduces $k_{\rm min}$ only by a factor $\sim 5$. So, in some sense, in our effort to go beyond the SPT  range, we have overshot. 

On the other hand, our results are quite robust, as they only depend on the assumption that dark matter is perfectly cold and can be described in terms of a displacement field with a well defined pdf. Therefore, any extra physical effect should be discussed in terms of the modifications it provides to these simple assumptions. 

Baryon feedback  modifies the dynamics of the CDM displacement field, but it does not modify its ``coldness'' appreciably, so it probably affects $k_{\rm min}$, but should not erase the asymptotic behavior. In this respect, notice that the lower curves in Fig.~\ref{Plot3D} are obtained by taking into account baryon feedback on the CDM distribution \cite{Mocz:2019uyd}, and we see that, at least at redshift $z=7$, the $1/k^3$ decay is there, as it is at $z=0$ for the CDM-only version of the  Illustris-TNG-100 simulation \cite{Springel:2017tpz}.  
Modifications to the CDM paradigm, as in ``warm" or ``fuzzy'' DM scenarios, do, on the other hand affect the coldness hypothesis, by introducing some sort of velocity dispersion, or pressure. This would smoothen caustics in Eulerian space, resulting in a exponential damping to the PS above a typical scale. In no scenario we can think of, however, and barring shot noise, a decay slower than $1/k^3$ appears to be possible, regardless of the dynamics and the dark matter properties, and this is probably the most model-independent statement we can draw at the moment. 

The question now is, clearly, how to decrease $k_{\rm min}$ and, eventually, bridge the gap between it and the $k_{\rm nl}$ scale, which limits from above the reach of SPT-like methods. The first step would be probably to match our approach with the halo model, which, as discussed in Sect.~\ref{halolink}, would mean some modification of the latter at very small scales. Then, this corrected halo model could be matched to SPT-like approaches (including UV counterterms)  along the lines discussed for example in \cite{Valageas:2010yw}, thus completing the ``ladder'' of theoretical approaches for the nonlinear LSS from the very large to the very small scales.

\section*{Acknowledgments}
We thank Marco Marinucci, Sabino Matarrese, and Matteo Viel for useful discussions, and  Philip Mocz and Volker Springel for providing us the simulation data of refs.~\cite{Mocz:2019uyd, Mocz:2019emo} and \cite{Springel:2017tpz}, respectively, used in Fig.~\ref{Plot3D}. MP acknowledges support from the European Union's Horizon 2020 research and innovation programme under the Marie Sklodowska-Curie grant agreements No 690575 and 674896. SC is supported by the National Science Foundation Graduate Research Fellowship (Grant No.~DGE 1106400) and by the UC Berkeley Theoretical Astrophysics Center Astronomy and Astrophysics Graduate Fellowship.

\appendix
\section{Nonperturbative correlation function in Zel'dovich dynamics}
\label{npZeld}

In this Appendix we discuss the correlation function in Zel'dovich dynamics, in which a lot can be seen analytically. Since the late-time field is proportional to the initial one (see Eq.~\re{zeldg}), it is gaussian, which greatly simplifies the computations of statistical averages. 
In this case, the correlation function is given by
\beqra
&& \!\!\!\!\!\!\!\!\!\!\!\!\!\!\!  1+\xi^Z(r)= \langle  \int_{-\infty}^{\infty} dq\,\delta_D\left(r-q-\Delta\psi\left(q\right) \right)\rangle_Z = \int_{-\infty}^{\infty}   dq \langle \delta_D\left(r-q-\Delta\psi\left(q\right) \right)\rangle_Z\,,\nonumber\\
&&\quad\;\,= \int_{-\infty}^{\infty}   dq \int \frac{{\cal D} \Delta\psi(q)}{\sqrt{2 \pi \sigma^2_{\Delta\psi}(q)}} e^{-\frac{1}{2} \frac{\Delta\psi(q)^2}{\sigma^2_{\Delta\psi}(q)}} \delta_D\left(r-q-\Delta\psi\left(q\right) \right)\,,\nonumber\\
&&\quad\;\,= \int_{-\infty}^{\infty}   \frac{dq }{\sqrt{2 \pi \sigma^2_{\Delta\psi}(q)}} e^{-\frac{1}{2} \frac{(q-r)^2}{\sigma^2_{\Delta\psi}(q)}}\,,
\label{xiZ}
\eeqra
where 
\beq
\sigma_{\Delta\psi}^2(q)\equiv \langle\Delta \psi(q)^2\rangle = q^2\int \frac{dp}{2\pi} \,W\left(\frac{p\,q}{2}\right)^2 P_{\rm{lin}}(p)\,,
\label{sigdeltapsi}
\eeq
with $W(x)\equiv \sin x/x$. We have  used the linear relation $\partial \psi(q)/\partial q=-\delta(q)$.
$\sigma_{\Delta\psi}^2(q)$ has the asymptotic behaviors,
\beqra
&&\sigma_{\Delta\psi}^2(q) \to q^2 \sigma_\delta^2\,, \;\;\;\mathrm{for}\;\;\;  q\ll \frac{\sigma_{\Delta\psi}^2(\infty)}{\sigma_\delta^2}\,,\nonumber\\
&& \sigma_{\Delta\psi}^2(q) \to \sigma_{\Delta\psi}^2(\infty)\,, \;\;\;\mathrm{for}\;\;\;  q\gg \frac{\sigma_{\Delta\psi}^2(\infty)}{\sigma_\delta^2}\,,
\label{sigmalims}
\eeqra
where
\beq
\sigma_{\Delta\psi}^2(\infty)= 2 \,\int\frac{dp}{2\pi}\,\frac{ P_{\rm{lin}}(p)}{p^2}\,\,,\qquad  \sigma_\delta^2=\int\frac{dp}{2\pi}\, P_{\rm{lin}}(p)\,.
\label{sigmas}
\eeq
Given the above behavior at small $q$, one sees that, as anticipated in Sect.~\ref{cfss}, Eq.~\re{xiZ} diverges logarithmically  in the $r \to 0$ limit.  
Indeed, when $r,\,|q|\ll  \sigma_{\Delta\psi}(\infty)/\sigma_\delta$, the integrand can be approximated as
\beqra
&&\int_{-A \sigma_{\Delta\psi}(\infty)/\sigma_\delta}^{A\sigma_{\Delta\psi}(\infty)/\sigma_\delta} \frac{dq }{\sqrt{2 \pi \sigma_\delta^2 q^2}} e^{-\frac{1}{2}\frac{(q-r)^2}{q^2\sigma_\delta^2}} = \frac{1}{\sqrt{2 \pi \sigma_\delta^2}}\int_{-A\sigma_{\Delta\psi}(\infty)/(r \sigma_\delta)}^{A\sigma_{\Delta\psi}(\infty)/(r \sigma_\delta)} \frac{dy}{y}e^{-\frac{1}{2}\frac{(y-1)^2}{y^2 \sigma_\delta^2}}\nonumber\\
&&=  \frac{e^{-\frac{1}{2\sigma_\delta^2}}}{\sqrt{2 \pi \sigma_\delta^2}} \log\left(\frac{\sigma_{\Delta\psi}(\infty)^2}{r^2 \sigma_\delta^2}\right) +\cdots\,,
\label{logdiv}
\eeqra
where $A\ll1$ and the dots indicate terms non-singular in the $r\to 0$ limit. The small-$r$ logarithm cannot be reproduced by SPT, as can be immediately seen by the non-perturbative $\sigma_\delta$ dependence of the prefactor. Moreover, we reiterate that the logarithmic divergence is not a peculiarity of the Zel'dovich dynamics, but we expect it to hold also for the exact dynamics, although with a different prefactor. This is confirmed by the results discussed in Sect.~\ref{PSS} on  the PS counterpart of this behavior of the correlation function. 

The SPT expression for the correlation function can be obtained by expanding the delta function in \re{2} and evaluating the terms of the series as 
\beqra
&& 
\!\!\!\!\!\!\! \!\!\!\!\!\!\! \!\!\!\!\! \!  \xi_{\rm SPT}( r) = \langle \sum_{n=1}^{\infty} \frac{(-1)^n}{n!} \frac{\partial^n}{\partial r^n}\Delta \psi(r)^n  \rangle_{Z} \,,\nonumber\\
&& \!\!\!\!\!\!\! \!\!\!\!\!\!\! \!\!\!\!\!\!\! \!\!\!\!\!\!\! \qquad\qquad \;=\sum_{n=1}^\infty \frac{1}{2^n \,n!} \frac{\partial^{2n}}{\partial r^{2n}} \left( \sigma_{\Delta\psi}^2(r)\right)^n\,\nonumber\\
&& \!\!\!\!\!\!\! \!\!\!\!\!\!\! \!\!\!\!\!\!\! \!\!\!\!\!\!\!\qquad\qquad \; =\frac{1}{2} \sigma^{2\,''}_{\Delta\psi}(r) + \frac{1}{4} \left( 3 (\sigma^{2\,''}_{\Delta\psi}(r))^2 + 4 \sigma^{2\,'}_{\Delta\psi }(r)\sigma^{2\,'''}_{\Delta\psi}(r) + \sigma^{2}_{\Delta\psi}(r)\sigma^{2\,^{''''}}_{\Delta\psi }(r)\right) +\cdots\,.\nonumber\\
\label{xiSPT}
\eeqra
As expected, even for  large $r$, where SPT is supposed to be a good tool, the above series does not converge. To see it, we first notice that at large $r$ we have
\beq
\sigma^2_{\Delta \psi}(r) \simeq \sigma^2_{\Delta \psi}(\infty) \left(1+O( \tilde r^{-2})\right)\,,
\eeq
where we have defined the dimensionless distance 
\beq
\tilde r \equiv \frac{r \sigma_\delta}{\sigma_{\Delta \psi}(\infty) }\,.
\eeq
Therefore, at large $\tilde r$,  
\beq
 \!\!\!\!\!\!\! \frac{1}{2^n \,n!} \frac{\partial^{2n}}{\partial r^{2n}} \left( \sigma_{\Delta\psi}^2(r)\right)^n\ \sim \frac{\,\sigma_\delta^{2n} }{2^n \,(n-1)!} \frac{\partial^{2n}}{\partial \tilde r^{2n}} \tilde r^{-2}
=\sigma_\delta^{2n}\, n (2n+1)!! \, \tilde r^{-2(n+1)}\,. \eeq
The factorially growing coefficients are typical of asymptotic series. 
Indeed, the condition for having the $(n+1)$-th term smaller then the $n$-th one is (for large $n$) 
\beq
\tilde r^2>2 n \sigma_\delta^2 \,,
\eeq
which shows that  for every finite $r$ ($\gg \sigma_{\Delta \psi}(\infty)/\sigma_\delta$) the series starts diverging for $n > O(r/ \sigma_{\Delta \psi}(\infty)) \gg 1/\sigma_\delta$.

Finally, we consider the case $r=0$, where a full analytical computation is possible provided we first regularize the correlation function in \re{xiZ} as
\beq
 \!\!\!\!\!\!   \!\!\!\!\!\!  \!\!\!\!\!\!  \!\!\!\!\!\!   \!\!\!\!\!\!   \bar\xi^Z(0)=\int\frac{d r}{\sqrt{2 \pi \sigma^2}} \, e^{-\frac{r^2}{2 \sigma^2}}\,\xi^Z(r) = -1+ \int_{-\infty}^{\infty}   \frac{dq }{\sqrt{2 \pi \left(\sigma^2 + \sigma^2_{\Delta\psi}(q)\right)^2}}
e^{- \frac{q^2}{2(\sigma^2+\sigma^2_{\Delta\psi}(q))}}\,.
\eeq
 Taking into account the asymptotic behaviors \re{sigmalims}, the integral can be approximated as
\beqra
&& \!\!\!\!\!\!   \!\!\!\!\!\!  \!\!\!\!\!\!   \!\!\!\!\!\!  \!\!\!\!\!\!    \bar\xi^Z(0) \simeq -1 +2\int_0^{\sigma/\sigma_\delta} \frac{dq}{\sqrt{2\pi(\sigma^2+q^2\sigma_\delta^2)}} e^{-\frac{q^2}{2 (\sigma^2+q^2\sigma_\delta^2)} }\nonumber\\
&&+2 \int_{\sigma/\sigma_\delta}^{\sigma_{\Delta\psi}(\infty)/\sigma_\delta} \frac{dq }{q \sqrt{2\pi \sigma_\delta^2 }}e^{-\frac{1}{2 \sigma_\delta^2} }\nonumber\\
&&+2  \int_{\sigma_{\Delta\psi}(\infty)/\sigma_\delta}^\infty  \frac{dq }{ \sqrt{2\pi \sigma_{\Delta\psi}^2(\infty) }}e^{-\frac{q^2}{2 \sigma_{\Delta\psi}^2(\infty)} }\,.
\eeqra

Notice that the integral at the second line is responsible for the logarithmic divergence as $\sigma\to 0$,  
\beq
\frac{1}{\sqrt{2\pi \,\sigma_\delta^2}} e^{-\frac{1}{2 \sigma_\delta^2}}\,\left( \log\left(\frac{\sigma^2_{\Delta\psi}(\infty)}{\sigma^2}\right)+C'\right)\,.
\eeq
The  integral at the first line can be written as
\beq
2\left(\int_0^\infty-\int_{1/\sigma_\delta}^\infty \right)\frac{dy}{\sqrt{2\pi(1+y^2\sigma_\delta^2)}} e^{-\frac{y^2}{2(1+y^2\sigma_\delta^2)}}\,,
 \eeq
 while the third integral gives $1-\mathrm{Erf}[1/\sqrt{2\sigma_\delta^2}]$.
Expanding in $\sigma_\delta^2$ the first line and sending $\sigma\to 0$, we get
 \beqra
 && \bar\xi^Z(0)\simeq \sum_{n=0}^\infty \frac{2^n \sigma_\delta^{2n}}{\sqrt{\pi}}\left(\Gamma\left(n+\frac{1}{2};0\right) - \Gamma\left(n+\frac{1}{2};\,\frac{1}{4\sigma_\delta^2}\right) \right)\nonumber\\
 && \qquad\quad+ \frac{1}{\sqrt{2\pi \,\sigma_\delta^2}} e^{-\frac{1}{2 \sigma_\delta^2}}\,\left( \log\left(\frac{\sigma^2_{\Delta\psi}(\infty)}{\sigma^2}\right)+C'\right)\nonumber\\
 && \qquad\quad-\mathrm{Erf}\left[\frac{1}{\sqrt{2\sigma_\delta^2}}\right]\,,
 \label{xi0}
 \eeqra
 where the incomplete gamma function is given by $\Gamma(n;x)=\int_x^\infty dt\, t^{n-1} e^{-t}$, and  $\Gamma(n+1/2;0)=  \sqrt{\pi} (2n-1)!!/2^n$.
 
 The first term reproduces the full SPT expansion, as it can be verified by setting $r=0$ in \re{xiSPT}. The coefficients are factorially growing, thereby confirming that it is an asymptotic series with zero radius of convergence. This pathologic behavior is cured by the nonperturbative series at the second term of the first line (notice that the two series coincide, term by term, in the $\sigma_\delta\to  \infty$ limit. The complete  expression is then just logarithmically divergent in the smoothing length $\sigma$, with a well defined (nonperturbative) coefficient, given at the second line. 

\section{Numerical details}
\label{numerics}
In this appendix we describe in detail the numerical procedure used to obtain Figs. 1 to 5. The algorithm to solve the equation of motion for the full dynamics has already been described in \cite{Pietroni:2018ebj}, and we recall it briefly.

Using as ``time'' variable  the logarithm of the scale factor,
\beq
\eta=\log\frac{a}{a_0}=-\log (1+z)\,,
\eeq
the equation of motion \re{fulleq} can be written as the system
\beqra
 &&\partial_\eta \psi(q,\eta) = \chi(q,\eta)\,,\nonumber\\
 &&\partial_\eta \chi(q,\eta) = -\frac{1}{2} \chi(q,\eta)+ \frac{3}{2} \sum_{i=1}^{N_s(x,\eta)} (-1)^{i+1} \psi(q_i(x,\eta),\eta)\,.
 \label{syst}
\eeqra
The initial condition is given at an early redshift in which we assume the linear theory growing mode, namely
\beq
\psi(q,\etain) = \chi(q,\etain) = \frac{v(q,\etain)}{\Hc(\etain)}  \,,
\eeq
where $v$ is the peculiar velocity.

The initial displacement field is represented as a Fourier series (see also \re{Fpsi}),
\beq
\psi(q,\etain) =\frac{2}{L} \sum_{n=1}^{N_p} c_n\, \cos\left(p_n \,q +\varphi_n\right)\,, \qquad\qquad\left(p_n\equiv \frac{2n\pi}{L}\right)\,,
\eeq
where the real amplitudes $c_n$ are randomly distributed following a Rayleigh distribution,
\beq
{\cal P}_R[\{c_n\}]=\Pi_n\, \frac{1}{2\pi}\, \frac{c_n}{\sigma_n^2} e^{-\frac{c_n^2}{2\sigma_n^2}}\,,
\eeq
with 
\beq
\frac{\sigma_n^2}{L}= \frac{1}{2}\frac{P_{\rm{lin}}(p_n;z_{\rm in})}{p_n^2},
\eeq
while the phases $\varphi_n$ are uniformly distributed on the interval $[0,2\pi)$. 

The solution of the above system of equations can then be computed by a straightforward algorithm, which requires just a few lines of code.
At each time-step, for each $x$ we identify the subset of Lagrangian points $\{q_i(x,\eta)\}$, containing all the real roots of the equation $x-q-\psi(q,\eta)=0$. Then, for each $q$, we compute the corresponding $x=q+\psi(q,\eta)$, and then  the increment of  $\psi(q,\eta)$, and $\chi(q,\eta)$, which involves, through the sum in \re{syst}, the previously identified subset $\{q_i(x,\eta)\}$ (which, of course, includes also $q$).

The plots presented in this paper have been obtained by setting dividing a line of $L=3000\, {\rm Mpc/h}$ into $N=24000$ grid points. We have also set $N_p=N$ and performed the time integration from $z_{\rm in}=99$ to $z=0$ in 500 time steps.

\section*{References}
\bibliographystyle{JHEP}
\bibliography{mybib}
\end{document}